\documentclass[10pt]{article}
\usepackage{amssymb,amsmath,amsfonts,amsthm,enumerate,color}
\usepackage[toc,page,title,titletoc,header]{appendix}
\usepackage{graphicx}
\usepackage{indentfirst}
\usepackage{multicol}
\usepackage{booktabs}

\numberwithin{equation}{section}

\numberwithin{equation}{section}

 \def\p{\partial} 
\def \Vh0{\stackrel{\circ}{V}_h} \def\to{\rightarrow}
   
\def\Om{\Omega}   
    
  \def\f{\frac}  

\newcommand{\lc}
{\mathrel{\raise2pt\hbox{${\mathop<\limits_{\raise1pt\hbox
{\mbox{$\sim$}}}}$}}}

\newcommand{\gc}
{\mathrel{\raise2pt\hbox{${\mathop>\limits_{\raise1pt\hbox{\mbox{$\sim$}}}}$}}}

\newcommand{\ec}
{\mathrel{\raise2pt\hbox{${\mathop=\limits_{\raise1pt\hbox{\mbox{$\sim$}}}}$}}}

\def\bb{\begin{equation}} \def\ee{\end{equation}}

\def\beqn{\begin{eqnarray}}  \def\eqn{\end{eqnarray}}

\def\beqnx{\begin{eqnarray*}} \def\eqnx{\end{eqnarray*}}

\def\bn{\begin{enumerate}} \def\en{\end{enumerate}}

\def\bd{\begin{description}} \def\ed{\end{description}}

\makeatletter

\newenvironment{figurehere}
  {\def\@captype{figure}}
  {}
\makeatother

\title{Direct Sampling Method for Diffusive Optical Tomography}

\begin{document}
\author{
Yat Tin Chow\footnote{Department of Mathematics, The Chinese University of Hong Kong, Shatin, Hong Kong. (ytchow@math.cuhk.edu.hk, kjliu@math.cuhk.edu.hk).}
\and Kazufumi Ito\footnote{Department of Mathematics and Center for Research in Scientific Computation, North Carolina State University, Raleigh, North Carolina (kito@unity.ncsu.edu).
}
\and Keji Liu\footnotemark[1]
\and Jun Zou\footnote{Department of Mathematics, Chinese University of Hong Kong, Shatin, N.T., Hong Kong.
The work of this author was substantially supported by Hong Kong RGC grants (projects 405513 and 404611).
(zou@math.cuhk.edu.hk).}
}

\date{}
\maketitle

\begin{abstract}
In this work, we are concerned with the diffusive optical tomography (DOT) problem in the case when only
one or two pairs of Cauchy data is available.
We propose a simple and efficient direct sampling method (DSM) to locate inhomogeneities inside a homogeneous background and solve the DOT problem in both full and limited aperture cases.
This new method is
easy to implement and less expensive computationally. 
Numerical experiments demonstrate its effectiveness and robustness against noise in the data.
This provides a new promising numerical strategy for the DOT problem.
\end{abstract}

\noindent { \footnotesize {\bf Mathematics Subject Classification
(MSC2000)}:
    35J67, 
    35R30, 
    65N21, 
    78A70, 
    78M25. 
}


\section{Introduction} \label{sec:intro}

Diffusive optical tomography (DOT) is a popular non-invasive imaging technique that measures the optical properties of a medium 
and creates images which show the distribution of absorption coefficient inside the body. 
In medical applications, such tomography is usually done in the near-infrared (NIR) spectral window. 
Chromophores in NIR window such as oxygenated and deoxygenated hemoglobin, water and lipid, are abundant in our body tissue, and a weighted sum of their contributions gives a different absorption coefficient \cite{thesis}. This provides us with detailed information about the concentration of various substances inside the tissue, thus revealing any pathological situation. 
In particular, it is known that cancerous tumors absorb light more than the surrounding tissues, therefore a very important medical application of DOT is the early diagnosis of breast cancer.  With the development of cost-efficient, compact and portable commercial instruments, we are now able to obtain accurate absorption measurements for bedside monitoring with a relatively low cost.  This propels the development of fast and accurate numerical algorithms to reconstruct the internal distribution of absorption coefficient so as to retrieve tissue properties in an accurate manner.  Medical applications of DOT include breast cancer imaging \cite{the62,the69,the102,the116,the119,the144,the165,the191,the194,the213,the222,the300}, brain
functional imaging \cite{the80}, stroke detection \cite{the23,the63,the133}, muscle functional studies \cite{the148,
the285},
photodynamic therapy \cite{the273,the282}, and radiation therapy monitoring \cite{the250}. More information about medical application can be found in \cite{thesis, paper}.


Over the past few decades, much effort has been made to develop efficient numerical algorithms to solve the DOT problem. Both the static and time-resolved models have been extensively studied.  For the time-resolved model, the problem is also known as photon migration, and it has been widely considered in literature, for example \cite{Timeresolvedpaper1,dynamicpaper1,dynamicpaper2}.  A popular approach to solve the DOT problem is to formulate a perturbation model, which comes from the linearization of the inverse problem \cite{FEMinvpaper1}. Another standard approach reduces the problem to an integral equation of the first kind, the kernel of which being generated by the Green's functions; see for instance \cite{Klibref4,Klibref2,Timeresolvedpaper1,Klibref29}. It is, however, well known that such an integral equation is ill-posed.  Therefore, the least-squares minimization with regularization is a usual technique to handle the problem.
Born type iterative method was introduced in \cite{bornpaper1},
based on the minimization of such a functional by regularized conjugate gradient method.
Nonetheless, a major drawback of these approaches to image small inclusions is the necessity to use a fine mesh,
resulting in a time-consuming process to invert highly ill-conditioned dense matrices.
Various other approaches have been investigated, e.g., the fast Fourier transform methods \cite{FFTpaper1}, the multigrid and Bayesian methods \cite{multigrid2,multigrid1}.  Effort has also been made to the three-dimensional reconstructions in optical tomography \cite{3Dpaper1,3Dpaper2}, as well as how to dramatically reduce the high computational complexity in handling large data set in solving the inverse problem \cite{largedatapaper1}. Other alternative directions have also been developed,
including the conversion of the original inverse problem to a well-posed system of elliptic partial differential equations \cite{Klibpaper2,Klibpaper1}, or to a Volterra-type integral differential equation \cite{Klibpaper3}.

In this work, we develop a novel direct sampling method (DSM) to solve
the DOT problem.  
Assume $\Omega\subset\mathbb{R}^d$ ($d=2, 3$) is an open bounded connected domain with a piecewise $C^2$ boundary, representing the absorption medium. Suppose that $\mu \in L^\infty(\Omega)$ is a non-negative function representing the absorption coefficient in $\Omega$, and $\mu_0$ is the absorption coefficient of the homogeneous background medium.  We shall often write the support of $\mu - \mu_0$ as $D$, which represents 
the inhomogeneous inclusions sitting inside $\Omega$.
Then the potential
$u\in H^1(\Omega)$ which represents the photon density field in the DOT model satisfies the following Neumann problem:
\begin{eqnarray}
\left\{
\begin{aligned}
-\Delta u+\mu u &=  0\quad \mathrm{in} \;\; \Omega\,,\\
\frac{\partial u}{\partial \nu}&=g\quad \mathrm{on}\;\; \partial\Omega\,,
\end{aligned}
\label{eqn11}
\right.
\end{eqnarray}
where $g \in H^{-\f{1}{2}}(\p \Omega)$ represents the surface flux along the boundary of $\Omega$.


Assume that we have only a partial measurement of the surface potential 
$u|_{\Gamma}$ at a relatively open subset of the boundary $\Gamma\subset\partial\Omega$\,, 
while the surface potential of 
$\partial\Omega \backslash \Gamma$ is unknown to us. Here and throughout the paper, we shall often write the surface potential over $\Gamma$ as $f$, i.e., $f := u|_{\Gamma}$.  Our inverse medium problem is then formulated as follows: given a single or a small number (e.g., $2$ to $3$) of pairs $(g, f)$ of the Cauchy data, we recover 
the locations and the number of inhomogeneities $D$ inside the homogeneous background absorption medium.

We first introduce a family of probing functions $\{\eta_x\}_{x\in\Omega}$ based on the monopole potential. Using these probing functions, we can then define an index function $I(x)$ which gives a clear image of the location of $D$ inside $\Omega$.  
From this index function, we can determine the number and locations of inclusions effectively.
We then move on to give an alternative characterization of the index function, which help enhance
the image quality as well as reduce computational complexity.
The new method is fast, reliable and cheap, so it can serve as an effective initialization for any iterative refinement algorithms,
such as the Born type iteraiton \cite{bornpaper1} and other more refined optimisation type methods
\cite{multigrid2,multigrid1}.
The new DSM is very different from the previous DSMs \cite{Ito,Ito13} developed
for inverse scattering problems
in the sense that the probing functions which we introduce here
may be regarded as the dual of the Green's function for the original forward problem \eqref{eqn11} under an appropriate choice of Sobolev scale.
Therefore our index function is in the dual form with a different Sobelov scaling
instead of the original $L^2$ inner product form adopted in the existing DSMs \cite{Ito,Ito13}.

The paper is organized as follows.
In section \ref{sec:DSM_sec}, we introduce the general philosophy of the DSM as well as defining the probing and index functions that we shall use in our DSM method for solving the DOT problem.
Then we give a representation of the index function in section \ref{sec:var}.
Explicit expressions of the probing function in some special domains are given in section \ref{sec:probing}, therefore giving us a clear picture on the performance of our newly proposed method in these special domains.
In section \ref{sec:scale}, we introduce a technique for improving the imaging as well as the computational complexity
of the index function. Numerical results are presented in section \ref{sec:num}, and
some concluding remarks are given in section \ref{sec:con}.

\section{The direct sampling method} \label{sec:DSM_sec}

\subsection{General principles of the DSM}\label{sec:philo}
In this subsection, we give a brief introduction of the general philosophy of the DSM.
Direct sampling type methods were studied in \cite{Ito, Ito13} for inverse scattering problems, based
on the well-known fact that the scattered field can be approximated by a finite sum of the fundamental solutions centered at the inhomogeneous scatterers.  With the help of a family of probing functions which is nearly orthogonal in some inner product space, an index function can be defined in such a way that it attains a large value inside the inhomogeneous scatterers but a small value outside.
This index function has been shown to
be a very effective method to reconstruct clustered scatterers
in two- and three-dimensional scattering media with a limited number of incident plane
waves. In this work, we shall develop a novel DSM for solving a very different inverse problem,
the DOT problem, based on some important probing functions, which
may be regarded as the dual of the Green's function for the original forward problem \eqref{eqn11}
under an appropriate choice of Sobolev scale. In this sense,
the new index function here is introduced in a dual form with a proper Sobolev scaling,
instead of the standard $L^2$ inner product used in the previous DSMs \cite{Ito, Ito13}.

In what follows, we derive a general framework of the DSM for the DOT problem.
We first consider the case when the absorption coefficient
of homogeneous background is non-zero ($\mu_0 \neq 0$).
We shall often write by $u_0$ the potential satisfying the diffusion equation
with homogeneous coefficient $\mu_0$, i.e.,
\beqn
-\Delta u_0+\mu_0 u_0 = 0 \quad\text{ in }\quad\Omega\, ; \quad  \frac{\partial u_0}{\partial \nu}=g \quad\text{ on }\quad\partial\Omega
\label{eqn111}
\eqn
and by $f_0$ the surface potential of $u_0$ along $\Gamma$, i.e. $f_0 := u_0|_{\Gamma}$.
We will also denote by $G_x$ the Green's function of the homogeneous diffusion equation:
\beqn
-\Delta G_x+\mu_0 G_x=\delta_x \quad \text{ in } \quad\Omega\,; \quad
\frac{\partial G_x}{\partial\nu} =0 \quad\text{ on }\quad \partial\Omega\,.
\label{Green}
\eqn
Now if $u$ is the solution to \eqref{eqn11}, it follows from the definition of $u_0$ that
\begin{equation}
-\Delta(u-u_0)+\mu_0(u-u_0)+(\mu-\mu_0)u=0 \quad\text{ in }\quad \Omega\,; \quad  \frac{\partial (u-u_0) }{\partial \nu}=0 \quad\text{ on }\quad\partial\Omega \,.
\label{diffuu0}
\end{equation}
Then by the Green's representation, we obtain the well-known Lippmann-Schwinger representation for the scattered potential $f-f_0$
of the DOT problem:
\beqn\label{eqn:property_bcts}
(f - f_0)(\xi) =\int_\Omega \,G_y(\xi) c(y)dy = \int_{D} \,G_y(\xi) c(y) \,dy \,  \quad \forall\,\xi \in \Gamma\,,
\eqn
where function $c$ is defined by 
\beqn
c(y):=- (\mu(y)-\mu_0)u(y)\, , \quad y \in \Om\,.
\label{defc}
\eqn
Using \eqref{eqn:property_bcts} and some numerical quadrature rule, we can approximate the scattered potential $f-f_0$ by a finite sum of fundamental solutions in the following form,
\begin{equation}\label{eqn:property_b}
(f-f_0)(\xi)\approx \sum_k a_k G_{x_k}(\xi)\,  \quad \forall\,\xi \in \Gamma\,,
\end{equation}
where $\{x_k\}$ are some quadrature points located inside $D$ and $\{a_k\}$ are some coefficients.

With the above approximation of the scattered field, we shall construct some index functions which may help locate the inhomogeneities of the medium.
In the subsequent discussion, we will often write $\nabla_{\Gamma}$, $\Delta_{\Gamma}$
and $d \sigma_{\Gamma}$ as the surface gradient, the surface Laplace operator
and the volumetric element on the surface $\Gamma$.
Consider the following duality product $\langle\, \cdot \, , \,  \cdot \, \rangle_s$ which is well-defined for the space $H^{2s}(\Gamma)
\times L^2(\Gamma) $ for some $s \in \mathbb{R}$ as follows:
\beqn
  \langle \phi , \psi \rangle_{s} :=  \int_{\Gamma} \, \left(-\Delta_{\Gamma}\right)^s \phi  \, \psi \, d \sigma_{\Gamma}\,  \quad \text{ for all } \phi \in H^{2s}(\Gamma)\,, \psi \in L^2(\Gamma) \,,
\label{Hs}
\eqn
then the space $L^2(\Gamma)$ can be considered as a subspace of the dual space of $H^{2s}(\Gamma)$ in the following sense:
\beqn
 \langle\, \cdot \, , \, \psi \, \rangle_s
 \in \left(H^{2s}(\Gamma)\right)^* \quad \forall \psi \in L^2(\Gamma) \,.
\label{duality}
\eqn
Assume that
$| \cdot |_Y$ is an algebraic function of semi-norms in $H^{2s}(\Gamma)$, and
we can select a set of probing functions $\{\eta_x\}_{x\in\Omega} \subset H^{2s}(\Gamma)$ such that
they are nearly orthogonal to the family $\{G_y \mid_{\Gamma} \}_{y\in\Omega}$
with respect to $\langle \cdot , \cdot  \rangle_s$ and $| \cdot |_Y$ in the following sense, namely for any $y \in \Omega$, the function
\beqn
x \mapsto K(x,y) := \frac{\langle \eta_x, G_y \rangle_s}{|\eta_x|_Y} \, , \quad x \in \Omega \,
\label{eqn11:K}
\eqn
attains maximum at $x=y$ and decays when $x$ moves away from $y$.
Under this assumption, we are now ready to define the following index function
\beqn
 I(x):=\frac{\langle \eta_x, f-f_0 \rangle_s}{|\eta_x|_Y}\,, \quad x \in \Omega \,.
 \label{indexgeneral}
\eqn
Directly substituting \eqref{eqn:property_b} into \eqref{indexgeneral}, we arrive at
\beqn
 I(x)=\frac{\langle \eta_x, f-f_0 \rangle_s }{|\eta_x|_Y} \approx \sum_ka_k\frac{\langle \eta_x, G_{x_k}\rangle_s}{|\eta_x|_Y} = \sum_ka_k K(x,x_k) \, ,
 \label{indexexpression}
\eqn
where $K$ is the function defined as in (\ref{eqn11:K}).

For the case when the absorption coefficient of the homogeneous background vanishes, namely
$\mu_0 = 0$, a similar representation of the index function $I$ can be obtained following the same argument as above.
In this case, the Green's function $G_x$ of the homogeneous diffusion equation is of the form:
\beqn
-\Delta G_x =\delta_x  \quad\text{ in } \quad\Omega\,; \quad
\frac{\partial G_x}{\partial\nu}=\f{1}{|\p\Om|} \quad\text{ on }\quad \partial\Omega\,; \quad \int_{\p\Om} G_x \, d \sigma = 0\,,
\label{Greenhomozero}
\eqn
whereas the incident potential $u_0$ satisfies
\beqn
-\Delta u_0 = 0\quad \text{ in }\quad\Omega\, ; \quad  \frac{\partial u_0}{\partial \nu}=g \quad\text{ on }\quad\partial\Omega \,; \quad \int_{\p\Om} u_0 \, d \sigma = 0\,.
\label{eqn111homozero}
\eqn
Similarly to the previous case, we can see that the difference $u -u_0$ satisfies \eqref{diffuu0} with $\mu_0 = 0$.
Hence we obtain the following Lippmann-Schwinger representation of $f-f_0$ by the Green's representation:
{
\small\beqn\label{eqn:property_bctshomozero}
(f - f_0)(\xi) =\int_\Omega \,G_y(\xi) c(y)dy - \f{1}{|\p\Om|} \int_{\p\Om} (f-f_0) d \sigma  = \int_{D} \,G_y(\xi) c(y) \,dy  - \f{1}{|\p\Om|} \int_{\p\Om} f \,d \sigma, \;\; \xi \in \Gamma\,,
\eqn
}where the function $c$ is defined as in \eqref{defc}.
Therefore, as in the previous case, we obtain the following approximation of the scattered potential $f-f_0$:
\begin{equation}\label{eqn:property_b_homozero}
(f-f_0)(\xi)\approx \sum_k a_k G_{x_k}(\xi) - \f{1}{|\p\Om|} \int_{\p\Om} f d \sigma \, , \quad \xi \in \Gamma\,,
\end{equation}
for some quadrature points $\{x_k\}$ located inside $D$ and some coefficients $\{a_k\}$.
Suppose that 
the family of probing functions $\{\eta_x\}_{x\in\Omega} \subset H^{2s}(\Gamma)$ have the following property:
\begin{equation}\label{eqn:property_X_homozero}
\langle \eta_x , \,\mathbf{1}\, \rangle_s = 0 \,  \quad \forall \;\; x \in \Omega \,,
\end{equation}
where the notation $\mathbf{1}$ stands for the constant function $\mathbf{1} (\xi) = 1 $ for all $\xi \in \Gamma$.
Substituting this expression into \eqref{indexgeneral}, we obtain the following for the index function $I(x)$:
\beqn
 I(x)=\frac{\langle \eta_x, f-f_0 \rangle_s }{|\eta_x|_Y} \approx \sum_ka_k\frac{\langle \eta_x, G_{x_k}\rangle_s }{|\eta_x|_Y} - \f{\int_{\p\Om} f \,d \sigma }{|\p\Om|} \frac{\langle \eta_x, \, \mathbf{1}\, \rangle_s }{|\eta_x|_Y} = \sum_ka_k K(x,x_k) \, ,
 \label{indexexpression2}
\eqn
where $K$ is the function defined as in (\ref{eqn11:K}).

From representations \eqref{indexexpression} and \eqref{indexexpression2} of the index function $I(x)$, we can see that the magnitude of $I(x)$ is relatively large inside $D$, while it is relatively small outside.  Therefore, if the magnitude of the index function $I(x)$ is relatively large at a point $x\in\Omega$\,, it is most likely that the point $x$ lies inside $D$\,. On the contrary,
if the magnitude of $I(x)$ is relatively small at $x$, it is then very likely that the point is at the homogeneous background. Therefore the index function provides us with an estimate of the location of $D$, and thus the number and locations of inclusions.

%

\subsection{Probing functions and index function}\label{sec:ProbIndex}
In this subsection, we introduce a family of probing functions and the index function which we will use in the DSM for the DOT problem. For this purpose, we first define, for any given $x \in \Omega$, a function $w_x$ satisfying the following system
\begin{equation}
-\Delta w_x+\mu_0 w_x=\delta_x\quad\text{ in }\quad \Omega\,; \quad w_x=0 \quad\text{ on }\quad \Gamma\,; \qquad \frac{\partial w_x}{\partial \nu} =0 \quad \text{ on } \quad\partial\Omega\backslash\Gamma\,.
\label{wx}
\end{equation}
For convenience, we may sometimes write the function $w_x(\,\cdot\,)$ as $w(x,\,\cdot\,)$\,.
Now, given any $x \in \mathbb{R}^d$, let $\Phi_x$ be the fundamental solution to the homogeneous diffusion equation, i.e.,
\beqn
-\Delta\Phi_x (y)+\mu_0\Phi_x(y)=\delta_x(y)\, , \quad y \in \mathbb{R}^d\
\label{phiphi}
\eqn
subject to the decay condition:
\beqn
\Phi_x(y) \rightarrow 0 \quad \text{ as }\quad |y| \rightarrow \infty \,.
\label{phidecay}
\eqn
Subtracting \eqref{wx} from \eqref{phiphi}, we readily obtain a representation of the function $w_x$ based on the fundamental solution $\Phi_x$:
\beqn
w_x=\Phi_x-\psi_x \, ,
\label{repsmooth}
\eqn
where $\psi_x$ is a function satisfying the following system:
\beqn
-\Delta\psi_x+\mu_0\psi_x =0 \quad\text{ in }\quad\Omega\,; \quad
\psi_x = \Phi_x  \quad\text{ on }\quad\Gamma\,; \quad
\frac{\partial \psi_x}{\partial \nu}=\frac{\partial \Phi_x}{\partial\nu} \quad\text{ on }\quad\partial\Omega\backslash\Gamma\,.
\label{psismooth}
\eqn
With these notions, we are now ready to define the family of probing functions for our purpose. For a fixed point $x\in\Omega$, we define the probing function $\eta_x$ as the surface flux of $w_x$ over $\Gamma$, i.e.,
\beqn
\eta_x (\xi) :=\frac{\partial w_x}{\partial\nu} (\xi) \,, \quad \xi \in \Gamma\,.
\label{probinggeneral}
\eqn
The probing function can be treated as the dual of the Green's function for the original forward problem \eqref{eqn11} along the boundary $\Gamma$ under a proper choice of Sobolev scaling.
Some mathematical justification and detailed analyses for this choice of probing functions will be provided in section \ref{sec:probing} for two special domains.

Now, we shall define an index function that will be used to solve the DOT problem.
To be more specific, we confine ourselves to the case with $s=1$, while similar definitions can be done
for the index $s\ne 1$. Then the duality product $\langle \cdot , \cdot \rangle_1 $ for the space $H^{2}(\Gamma)
\times L^2(\Gamma) $ can be explicitly given as the following: 
\beqn
  \langle \phi , \psi \rangle_{1} := - \int_{\Gamma} \, \Delta_{\Gamma} \phi  \, \psi \, d \sigma_{\Gamma}\, , \quad \text{ for all } \phi \in H^2(\Gamma)\,, \psi \in L^2(\Gamma) \,.
\label{H1}
\eqn
As a remark, if $\Gamma$ is a closed surface and function $\psi\in H^1(\Gamma)$, then the above duality product actually equals to the well-known $H^1(\Gamma)$ semi-inner product, i.e., $\langle \phi , \psi \rangle_{1} = ( \phi , \psi )_{H^1(\Gamma)}$, where
\beqn
 ( \phi , \psi )_{H^1(\Gamma)} := \int_{\Gamma} \nabla_{\Gamma} \phi  \cdot \nabla_{\Gamma} \psi d \sigma_{\Gamma}\,, \quad \text{ for all } \phi, \psi \in H^1(\Gamma)\,.
\label{H12}
\eqn
One choice of the algebraic function $| \cdot |_Y$ in \eqref{eqn11:K} of semi-norms
can be ${|\, \cdot \,|^{1/2}_{H^1(\Gamma)}|\, \cdot \,|^{3/4}_{H^0(\Gamma)}}$, with $|\,\cdot \,|_{H^0(\Gamma)} := |\,\cdot \,|_{L^2(\Gamma)}$ and $|\,\cdot \,|_{H^1(\Gamma)} :=
( \, \cdot \,, \,\cdot\, )^{1/2}_{H^1(\Gamma)}$.
Substituting these expressions into \eqref{eqn11:K} and \eqref{indexgeneral}, we have the following expressions of the kernel $K$ and the index function $I$:
\beqn
K(x,y) &:=& \frac{\langle \eta_x, G_y \rangle_{1}} {|\eta_x|^{\f{1}{2}}_{H^1(\Gamma)}|\eta_x|^{\f{3}{4}}_{H^0(\Gamma)}} \,  \quad
\forall\, x,y \in \Omega \,, \label{eqn:Ksub} \\[3mm]
I(x)&:=&
\f{ \langle \eta_x, f - f_0 \rangle_{1} } {|\eta_x|^{\f{1}{2}}_{H^1(\Gamma)}|\eta_x|^{\f{3}{4}}_{H^0(\Gamma)}}
\, \quad\forall\,x \in \Omega \,.
\label{eqn:inner}
\eqn
%
%
We will show in section \ref{sec:probing} that,
for two special domains $\Omega$, namely the circular domain and the rectangular domain,
this index function has the desired property as stated in section \ref{sec:philo}: its magnitude is relatively large inside $D$ but relatively small outside.
Using this property, the inhomogeneities of the absorption medium $\Omega$ can be clearly located.
Unlike the most conventional methods for the DOT problem,
which use the regularized least-squares methods or iterative methods, this method is computationally inexpensive,
since it involves only the evaluation of a duality product $\langle \, \cdot \, , \, \cdot \, \rangle_1$, which is an integral over
the measurement surface $\Gamma$. Moreover, the noise which enters into the boundary data is averaged and smoothed by integration, and no matrix inversion is required.
In section \ref{sec:scale}, some technique will be introduced to further improve the quality of imaging as well as sufficiently reducing the computational cost by avoiding the calculation of the family of probing functions.
Therefore this method is expected to be efficient computationally and robust against the noise in the data,
which are demonstrated by numerical examples in section \ref{sec:num}.

\section{A representation of the index function} \label{sec:var}

In this section we try to give a representation of the $\langle \, \cdot \, , \, \cdot \, \rangle_{s}$ duality product between the probing functions $\eta_x$ and the measurement boundary data $f - f_0$, and therefore the index function. This representation provides a more detailed understanding of the behaviour of the index function.

We first consider the case with $s = 0$, then the duality product $\langle \,\cdot\,,\,\cdot\,\rangle_0$ is
the standard $L^2(\Gamma)$ inner product.
Given a point $x \in \Omega$\,, from \eqref{diffuu0}, \eqref{wx}, \eqref{probinggeneral} and the Green's identity, we have the following representation:
\beqn
( \eta_x, f - f_0  )_{0}  &=& \int_{\Gamma} \eta_x (u-u_0) \, d \sigma_{\Gamma} \notag\\
&=& \int_{\Gamma} (u-u_0) \frac{\partial w_x}{\partial\nu} \, d \sigma_{\Gamma} \notag\\
&=& \int_{\partial \Omega} \left( (u-u_0) \frac{\partial w_x}{\partial\nu} - w_x \frac{\partial  (u-u_0)}{\partial\nu} \right) \, d \sigma \notag\\
&=&
\int_\Omega \left( (u-u_0) \Delta w_x - w_x \Delta  (u-u_0) \right) \, d y \notag\\
&=&
- (u-u_0)(x) - \int_{D} (\mu-\mu_0)u w_x \, d y \notag\\
&\approx& - (u-u_0)(x) - \sum_k \lambda_k \left(\mu(y_k)-\mu_0\right) u(y_k)w_x(y_k)\,,
\eqn
for some quadrature points $y_k$ sitting inside in the inhomogeneous support $D$ and some quadrature weights $\lambda_k$ of the respective quadrature rule. Therefore we can find that if $x$ is near the point $y_k$ for some $k$, then the magnitude of the duality product $\langle \eta_x, f - f_0  \rangle_{0}$ is relatively large, otherwise it would be relatively small.


Next, we consider the case with $s = 1$ and the
duality product $ \langle \cdot , \cdot \rangle_{1}$ as defined in \eqref{H1}.
We consider the rectangular domain
$\Omega:=(0,h)\times (-L,L)$ for some $h,L \in \mathbb{R}$ and $\Gamma = \{0,h\}\times (-L,L) \subset \p \Omega$.
The subsequent derivations apply naturally to $d$-dimensional rectangular domains.

From \eqref{diffuu0}, we can see that the function $\p^2_{x_2 x_2} (u-u_0)$ satisfies the following equation:
\begin{equation}
-\Delta\big( \p^2_{x_2 x_2}(u-u_0) \big) +\mu_0 \p^2_{x_2 x_2}(u-u_0) +  \p^2_{x_2 x_2}[(\mu-\mu_0) u]  =0 \quad \text{ in } \Omega \,.
\label{duminusu0}
\end{equation}
Now assuming $ u-u_0 \in H^{2}(\Omega)$, 
then using that $\nabla_{\Gamma} = \p_{x_2}$ and $\Delta_{\Gamma} = \p^2_{x_2 x_2}$ on $\Gamma$,
we have  for any $x \in \Omega$ by the Green's identity on $\Gamma$ the following duality product
of the probing function $\eta_x$ and the scattered potential $f -f_0$:
\beqn
\langle \eta_x, f - f_0  \rangle_{1} &=& - \int_{\Gamma}
\Delta_{\Gamma} \eta_x  (u-u_0) \, d \sigma_{\Gamma} \notag \\
&=& \int_{\Gamma}
\nabla_{\Gamma} (u-u_0) \cdot \nabla_{\Gamma} \eta_x \, d \sigma_{\Gamma} -  (u - u_0) (0,L) \f{\p \eta_x }{ \p x_2 } (0,L) + (u - u_0) (0,-L) \f{\p \eta_x }{ \p x_2 } (0,-L) \notag\\
& & -   (u - u_0) (h,L)  \f{\p \eta_x }{ \p x_2 } (h,L) + (u - u_0) (h,-L) \f{\p \eta_x }{ \p x_2 } (h,-L) \,.
\eqn
However, from \eqref{wx} and the fact that $(0,1)$ is normal to the surface $\p\Om \backslash \Gamma$,
we have that
\beqn
\f{\p \eta_x }{ \p x_2 } (0,L) =  \f{\p^2 w_x }{\p x_2 \p x_1} (0,L) =  0 \,.
\eqn
Similarly we can derive
\beqn
 \f{\p \eta_x }{ \p x_2 } (0,L) = \f{\p \eta_x }{ \p x_2 }  (0,-L) =  \f{\p \eta_x }{ \p x_2 } (h,L) = \f{\p \eta_x }{ \p x_2 }  (h,-L) = 0 \,.
 \label{zero110}
\eqn
Therefore we have
\beqn
\langle \eta_x, f - f_0  \rangle_{1}
&=& \int_{\Gamma}
\nabla_{\Gamma} (u-u_0) \cdot \nabla_{\Gamma} \eta_x \, d \sigma_{\Gamma}  \notag\\
&=& \int_{\Gamma}
\p_{x_2}(u-u_0) \p_{x_2}\eta_x \, d \sigma_{\Gamma} \notag\\
&=& - \int_{\Gamma}
\p^2_{x_2 x_2}  (u-u_0)  \eta_x \, d \sigma_{\Gamma}  +  \eta_x(0,L) \f{\p (u - u_0)}{ \p x_2 } (0,L) - \eta_x(0,-L) \f{\p (u - u_0)}{ \p x_2 } (0,-L) \notag\\
& & +  \eta_x(h,L) \f{\p (u - u_0)}{ \p x_2 } (h,L) - \eta_x(h,-L) \f{\p (u - u_0)}{ \p x_2 } (h,-L) \,.
\label{derive}
\eqn
However, again, from the fact that $(0,1)$ is normal to the surface $\p\Om \backslash \Gamma$, we get from \eqref{eqn11} and \eqref{eqn111} that
\beqn
\f{\p (u - u_0)}{ \p x_2 } (0,L) = \f{\p u }{ \p x_2 } (0,L) - \f{\p u_0}{ \p x_2 } (0,L)= g(0,L)- g(0,L) = 0 \,.
\eqn
Similarly we can derive
\beqn
 \f{\p (u - u_0)}{ \p x_2 } (0,L) = \f{\p (u - u_0)}{ \p x_2 } (0,-L) =  \f{\p (u - u_0)}{ \p x_2 } (h,L) = \f{\p (u - u_0)}{ \p x_2 } (h,-L) = 0 \,.
 \label{zero11}
\eqn
Using \eqref{derive}, \eqref{wx}, \eqref{probinggeneral} and the Green's identity, we obtain
\beqn
\langle \eta_x, f - f_0  \rangle_{1}
&=& - \int_{\Gamma} \p^2_{x_2 x_2}  (u-u_0)  \eta_x  \, d \sigma_{\Gamma}  \notag\\
&=& - \int_{\Gamma} \p^2_{x_2 x_2}  (u-u_0)  \frac{\partial w_x}{\partial\nu}   \, d \sigma_{\Gamma}  \notag\\
&=& - \int_{\Gamma} \left( \p^2_{x_2 x_2}  (u-u_0)  \frac{\partial w_x}{\partial\nu} -  w_x  \frac{\partial}{\partial\nu} \left( \p^2_{x_2 x_2}  (u-u_0) \right) \right) \, d \sigma_{\Gamma} \notag\\
&=& - \int_{\p \Om} \left( \p^2_{x_2 x_2}  (u-u_0)  \frac{\partial w_x}{\partial\nu} -  w_x  \frac{\partial}{\partial\nu} \left( \p^2_{x_2 x_2}  (u-u_0) \right) \right) \, d \sigma  \notag\\
&=& - \int_{\Om} \left( \p^2_{x_2 x_2}  (u-u_0) \Delta  w_x -  w_x  \Delta \left( \p^2_{x_2 x_2}  (u-u_0) \right) \right) \, d y + \varphi (x)\notag\\
&=&
\p^2_{x_2 x_2}(u-u_0)(x) + \int_{D} \p^2_{x_2 x_2}\left[(\mu-\mu_0) u\right] w_x \, dy \\
&\approx&  \p^2_{x_2 x_2}(u-u_0)(x) + \sum_k \lambda_k  \p^2_{x_2 x_2}\left[(\mu-\mu_0) u\right](y_k)w_x(y_k) \,
\eqn
for some quadrature points $y_k$ sitting inside in the inhomogeneous support $D$ and some quadrature weights $\lambda_k$ of the respective quadrature rule. We can now observe that $\langle \eta_x, f - f_0  \rangle_{1}$ is relatively large if $x$ is near $y_k$ for some $k$.

\section{Probing and index functions on two special domains} \label{sec:probing}

Following the framework introduced in section \ref{sec:philo}, in order to justify the DSM for a given domain $\Omega$,
we should show that
the family of probing functions defined as in \eqref{probinggeneral} satisfies two properties: the first
one is \eqref{eqn:property_X_homozero}, which will then validate \eqref{indexexpression2};
the second one is that they are nearly orthogonal to the family $\{G_y \mid_{\Gamma} \}_{y\in\Omega}$
with respect to $\langle \cdot , \cdot  \rangle_s$ and $| \cdot |_Y$ as stated in Subsection \ref{sec:philo}.
It is not easy to deduce these properties for a general domain $\Omega$.
In this section, we will calculate explicitly the probing function defined in \eqref{probinggeneral} and the kernel as in \eqref{eqn:Ksub} for two special domains $\Omega$, or two most frequently used sampling
domains: a circular domain and a rectangular domain, which help us
establish the two necessary properties to justify the DSM.

\subsection{The circular domain}
Let $\Omega$ be the the circular domain $B_1(0)$, i.e., the open ball of radius $1$ centered at origin in $\mathbb{R}^2$.
Consider the boundary measurement on $\Gamma = \partial B_1(0) = \mathbb{S}^1$, and
the case with $\mu_0=0$ first.
Then for $x \in \Omega$, we can see from \eqref{wx} that the function $w_x$ is actually the Green's function for the unit disk with the Dirichlet boundary condition:
\beqn
w(x,y) = \f{1}{2\pi}\left( \, \log |x-y| - \log \bigg|\f{x}{|x|} - |x|y \bigg| \, \right)\,, \quad y \in B_1(0) \,.
\eqn
Therefore the probing function $\eta_x$ defined in \eqref{probinggeneral} is the Poisson kernel with
the explicit form:
\beqn
    \eta_x (y) = \f{1 - |x|^2}{2\pi |x-y|^2}\,, \quad y \in \mathbb{S}^1 \,.
\eqn

Next, we shall calculate the kernel \eqref{eqn:Ksub} for the better understanding of the index function $I$ defined as in \eqref{eqn:inner}.  For the notational sake, we shall write
the Fourier coefficient of a function $\phi \in L^2(\mathbb{S}^1)$ as
\beqn
[\mathfrak{F} ( \phi ) ](n) := \f{1}{2\pi} \int_0^{2\pi} \phi(\theta) e^{- i n \theta} \, d \theta\,, \quad n \in \mathbb{Z}\,.
\eqn
Now for any $\phi,\psi \in H^1(\mathbb{S}^1)$, their duality product $\langle \, \cdot \, , \, \cdot \, \rangle_1$
is equal to the $H^1$ semi-inner product and can be expressed in the Fourier coefficients of $\phi$ and $\psi$ as
\beqn
\f{1}{2\pi} \langle \phi,\psi \rangle_{1} =  \sum_{n \in \mathbb{Z}} n^2 \overline{ [\mathfrak{F} ( \phi ) ](n)} [\mathfrak{F} ( \psi ) ](n) \, ,
\label{innerproducth1}
\eqn
From the fact that both the functions $G_z$ and $\eta_x$ are in $H^1(\mathbb{S}^1)$,
we can obtain an explicit expression of the kernel $K(x,z)$ for $x,z\in B_1(0)$ by calculating the Fourier expansions of $G_z$ and $\eta_x$.
For this purpose, we consider for a given function $g$ the following Neumann problem:
\beqn
- \Delta v  = 0 \quad \text{ in } \quad B_1(0) \,; \quad
\f{\p v}{\p \nu} = g \quad \text{ on } \quad  \mathbb{S}^1 \,; \quad
\int_{\mathbb{S}^1} v \, d \sigma = 0\,.
\eqn
On one hand, from the Green's representation formula, function $v$ can be expressed by
\beqn
v (z) =  \int_{\mathbb{S}^1} g (y) G_z (y) d \sigma(y)  - \f{1}{2 \pi} \int_{\mathbb{S}^1} v(y) \, d \sigma (y) =  \int_{\mathbb{S}^1} g (y) G_z (y) d \sigma(y)  \,, \quad y \in \mathbb{S}^1  \,,
    \label{fourier1}
\eqn
where $G_z$ is defined as in \eqref{Greenhomozero}. On the other hand, by a separation of variables, we can explicitly calculate the Fourier expansion of function $v$ as follows: 
\beqn
    v (z) =  \f{1}{2\pi}\sum_{n \in \mathbb{Z} \backslash \{0\}}  \f{r_z^{|n|} e^{i n \theta_z }}{|n|} \int_{0}^{2\pi} g(\theta_y) e^{- i n \theta_y} d \theta_y  =  \f{1}{2\pi} \int_{\mathbb{S}^1} g (y) \left(\sum_{n \in \mathbb{Z} \backslash \{0\}}  \f{r_z^{|n|}}{|n|}e^{i n (\theta_z - \theta_y)} \right) d y\,,
    \label{fourier2}
\eqn
where $z = (r_z,\theta_z)$ is in the polar coordinate.  Then comparing \eqref{fourier2} with \eqref{fourier1}, we obtain the Fourier expansion of $G_z$ as
\beqn
    G_z(y) =  \f{1}{2\pi}  \sum_{n \in \mathbb{Z} \backslash \{0\}}  \f{r_z^{|n|}}{|n|}e^{i n (\theta_z - \theta_y)} =  \f{1}{2\pi}  \sum_{n \in \mathbb{Z} \backslash \{0\}}  \f{r_z^{|n|}e^{- i n \theta_z}}{|n|}e^{i n \theta_y} \,, \quad y \in \mathbb{S}^1  \,.
    \label{coeff1}
\eqn
With a similar technique, we try to obtain the Fourier expansions of $w_x$. For a given function $h$, consider
the following Dirichlet problem:
\beqnx
- \Delta v  = 0 \quad\text{ in }\quad B_1(0) \,; \quad \quad
v = h\quad\text{ on } \quad\mathbb{S}^1 \,.
\eqnx
Then, by the definition of $w_x$ in \eqref{probinggeneral} and the Green's representation formula, we can see that the function $v$ can be expressed as
\beqn
v(x) =  \int_{\mathbb{S}^1} h (y) \f{\p w_x}{\p \nu}(y) \, d \sigma(y) \,, \quad x \in \mathbb{S}^1  \,.
\label{fourier3}
\eqn
On the other hand, we can obtain the following Fourier expansion of function $v$ by a separation of variables:
\beqn
v(x) = \f{1}{2\pi} \sum_{n \in \mathbb{Z} } r_x^{|n|}e^{i n \theta_x } \int_{0}^{2\pi} g(\theta_y) e^{- i n \theta_y} d \theta_y  = \f{1}{2\pi}  \int_{\mathbb{S}^1} g (y) \left( \sum_{n \in \mathbb{Z} }  r_x^{|n|}e^{i n (\theta_x - \theta_y)} \right) d y \,, \quad x \in \mathbb{S}^1  \,,
\label{fourier4}
\eqn
where $x = (r_x,\theta_x)$ is in the polar coordinate. Hence, we can compare \eqref{fourier4} with \eqref{fourier3} to obtain the following Fourier expansion for the probing function defined as in \eqref{probinggeneral}:
\beqn
\eta_x(y) = \f{\p w_x}{\p \nu}(y) =  \f{1}{2\pi} \sum_{n \in \mathbb{Z} } r_x^{|n|} e^{i n (\theta_x - \theta_y)} = \f{1}{2\pi} \sum_{n \in \mathbb{Z} } r_x^{|n|} e^{-i n \theta_x} e^{i n \theta_y}\,, \quad y \in \mathbb{S}^1  \,.
\label{coeff2}
\eqn
From \eqref{coeff1} and \eqref{coeff2}, we get the Fourier coefficients of $G_z$ and $\eta_x$:
\beqn
[\mathfrak{F} ( G_z ) ](n) =  \f{1}{2\pi} \f{r_z^{|n|}e^{- i n \theta_z}}{|n|} \,  \quad \forall \;\;n \in \mathbb{Z}\backslash\{0\}\, ; \quad \quad [\mathfrak{F} ( G_z ) ](0) =  0 \,, \label{coeffsln1}
\eqn
and
\beqn
 [\mathfrak{F} ( \eta_x ) ](n) = \f{1}{2\pi} r_x^{|n|} e^{-i n \theta_x} \,  \quad \forall\;\; n \in \mathbb{Z}\,. \label{coeffsln2}
\eqn
Substituting the above two expressions into \eqref{innerproducth1}, we readily obtain for all $x,z\in B_1(0)$,
\beqn
\langle \eta_x, G_z  \rangle_1 &=&  2\pi \sum_{n \in \mathbb{Z}} n^2 \overline{  [\mathfrak{F} ( \eta_x ) ](n)} [\mathfrak{F} ( G_z ) ](n) \notag\\
&=&
 \f{1}{2\pi} \sum_{n \in \mathbb{Z} \backslash \{0\}} |n| r_z^{|n|} r_x^{|n|}  e^{i n (\theta_x - \theta_z)}
=
 \f{1}{\pi} \text{Re} \Big( \sum_{n =1}^\infty n r_z^{n} r_x^{n}  e^{i n (\theta_x - \theta_z)} \Big) \, \notag\\
&=&
 \f{1}{\pi} \text{Re} \Big( r_z r_x e^{i (\theta_x - \theta_z)} \sum_{n =1}^\infty  n r_z^{n-1} r_x^{n-1}  e^{i (n-1) (\theta_x - \theta_z)} \Big)\,,
\eqn
which gives
\beqn
\langle \eta_x, G_z  \rangle_1 =
 \f{1}{\pi} \text{Re} \Big( \f{r_z r_x e^{i (\theta_x - \theta_z)} }{\left( 1 - r_z r_x  e^{i (\theta_x - \theta_z)}\right)^2 } \Big)
=
 \f{1}{\pi} \f{  r_z r_x \cos (\theta_x - \theta_z)(1 + r_z^2 r_x^2 ) - 2 r_z^2 r_x^2 }{ (1 -2 r_z r_x  \cos (\theta_x - \theta_z) +  r_z^2 r_x^2 )^2 }\,.
 \label{ans1}
\eqn
An interesting point to observe is that the duality product $\langle \eta_x, G_z  \rangle_1$ is actually symmetric about $x$ and $z$.  Moreover, the $L^2$-norm of $\eta_x$ can be obtained from the Plancherel identity as
\beqn
|\eta_x|^2_{H^0(\mathbb{S}^1)} = 2 \pi \sum_{n \in \mathbb{Z}} \Big| [\mathfrak{F} ( \eta_x ) ](n) \Big|^2
= \f{1}{2 \pi} \sum_{n \in \mathbb{Z}} r_x^{2|n|}
= \f{1}{2\pi}\left( \f{2 r_x^2 }{1-r_x^2} + 1 \right) = \f{1}{2\pi}\left( \f{1 + r_x^2 }{1-r_x^2} \right)\,,
 \label{ans2}
\eqn
whereas the $H^1$ semi-norm of $\eta_x$ can be calculated explicitly by \eqref{innerproducth1} as
\beqn
|\eta_x|_{H^1(\mathbb{S}^1)}^2 &=&
2 \pi \sum_{n \in \mathbb{Z}} n^2 \Big| [\mathfrak{F} ( \eta_x ) ](n) \Big|^2
=
\f{1}{2 \pi} \sum_{n \in \mathbb{Z}} n^2 r_x^{2|n|} \notag\\
&=&
\f{1}{ \pi} \sum_{n =0}^\infty \big( (n+2)(n+1) - 3(n+1) + 1 \big) r_x^{2n} \notag\\
&=&
\f{1}{ \pi}  \f{2 - 3(1-r_x^2)+ (1-r_x^2)^2}{(1-r_x^2)^3}
=
\f{1}{ \pi}\f{ r_x^2 (r_x^2+1) }{(1-r_x^2)^{3}}\,.
 \label{ans3}
\eqn

\begin{figurehere}
     \begin{center}
     \vskip -0.18truecm
           \scalebox{0.35}{\includegraphics{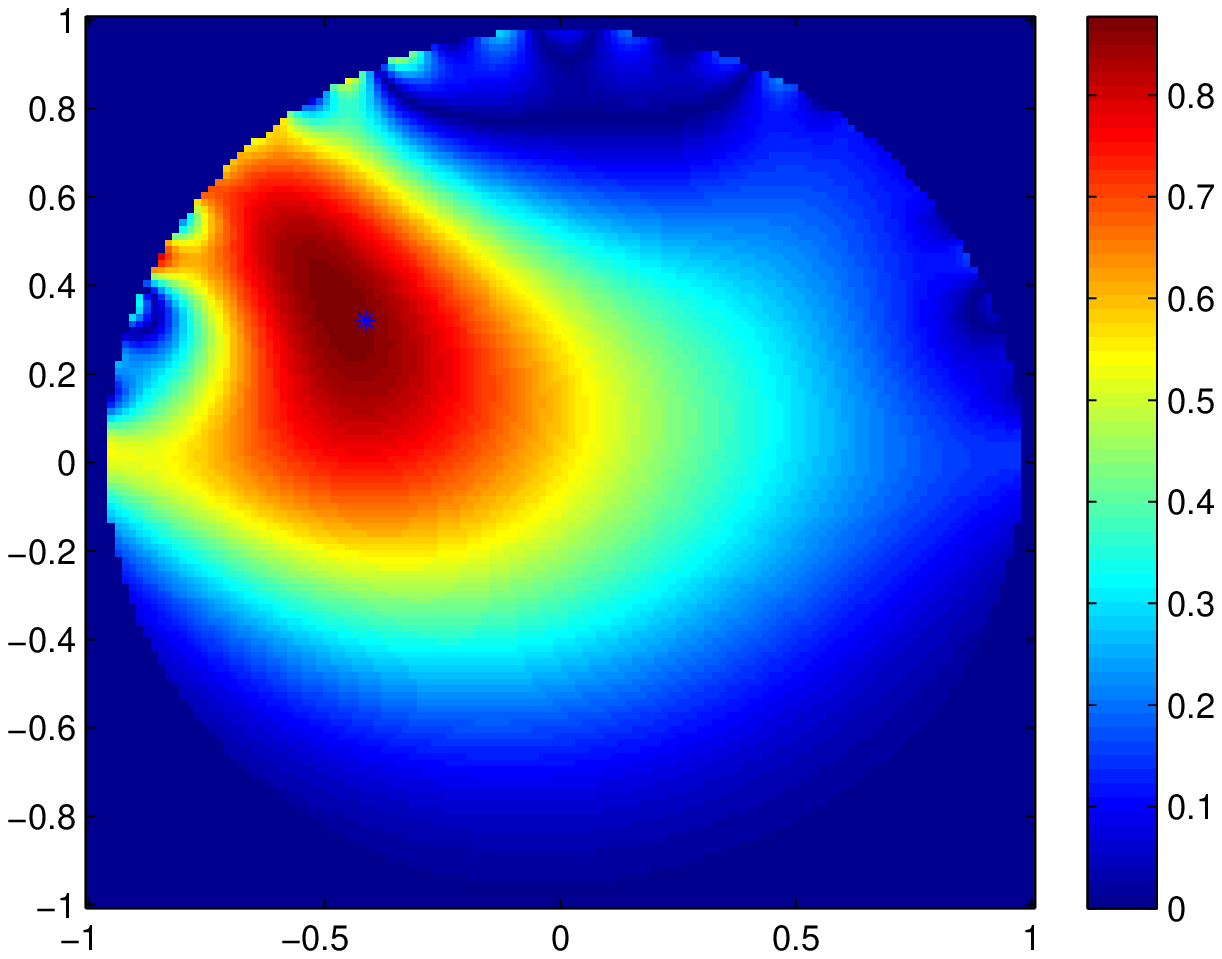}}\\
            \vskip -0.68truecm
     \caption{\small Kernel $K(\,\cdot\,,z)$ for a circular domain with $z = (-0.41, 0.32)$ marked as a blue star.} \label{kernal_circular_half}
     \end{center}
 \end{figurehere}

Substituting \eqref{ans1}-\eqref{ans3} into \eqref{eqn:Ksub}, we obtain the explicit expression
of kernel $K$ for all $x,z\in B_1(0)$:
\beqn
K(x,z) = \frac{\langle \eta_x, G_z \rangle_1 }{|\eta_x|^{\f{1}{2}}_{H^1(\Gamma)}|\eta_x|^{\f{3}{4}}_{H^0(\Gamma)}} =
C \left( \f{1-r_x^2}{1 + r_x^2} \right)^{\f{3}{8}}
\left( \f{ (1-r_x^2)^{3} }{r_x^2 (r_x^2+1)} \right)^{\f{1}{4}}
\left( \f{  r_z r_x \cos (\theta_x - \theta_z)(1 + r_z^2 r_x^2 ) - 2 r_z^2 r_x^2 }{ (1 -2 r_z r_x  \cos (\theta_x - \theta_z) +  r_z^2 r_x^2 )^2} \right). \quad
\eqn

From this expression, we can observe that the kernel $K(x,z)$ has a relatively large value when $x \approx z$. 
This validates the DSM for the cirular domain $B_1(0)$.
We remark that we may not be able to obtain full boundary measurements in practice.  Instead, we usually encounter the case where we only have measurements along the upper half of the boundary, i.e., when $\Gamma = \{ x \in \mathbb{S}^1 \, ; \; x_2 >0 \}$.
Figure \ref{kernal_circular_half} shows the values of distribution of kernel $K(x,z)$ with this $\Gamma$
and $z = (-0.41, 0.32)$ for $x \in B_1(0)$.
This figure indeed shows the distribution kernel $K(x,z)$ has a relatively large value when $x \approx z$,
and justifies the DSM for the circular domain with partial measurements.

Now we consider the case with $\mu_0\neq 0$. For a given $h$, consider the following Dirichlet problem:
\beqnx
- \Delta v + \mu_0 v = 0 \quad\text{ in }\quad B_1(0) \,; \quad \quad
v = h\quad\text{ on } \quad\mathbb{S}^1 \,.
\eqnx
Then, from the definition of $w_x$ as stated in \eqref{probinggeneral} and the Green's representation formula, we can see that the function $v$ can be expressed as
\beqn
v(x) =  \int_{\mathbb{S}^1} h (y) \f{\p w_x}{\p \nu}(y) \, d \sigma(y) \,, \quad x \in \mathbb{S}^1  \,.
\label{fourier5}
\eqn
However, by a separation of variables, we can obtain the following expansion of $v$ for $x \in \mathbb{S}^1$,
\beqn
v(x) &=& \f{1}{2\pi} \sum_{n \in \mathbb{Z} } \frac{J_{n}(i \sqrt{\mu_0} r_x )}{J_n(i\sqrt{\mu_0})}e^{i n \theta_x } \int_{0}^{2\pi} g(\theta_y) e^{- i n \theta_y} d \theta_y \notag\\
  &=& \f{1}{2\pi}  \int_{\mathbb{S}^1} g (y) \left( \sum_{n \in \mathbb{Z} }  \frac{J_{n}(i \sqrt{\mu_0} r_x )}{J_n(i\sqrt{\mu_0})} e^{i n (\theta_x - \theta_y)} \right) d y \,,
\label{fourier6}
\eqn
where $J_n$ are the Bessel functions of first kind of order $n$ and $x = (r_x,\theta_x)$ is in its polar coordinate.  Comparing \eqref{fourier5} with \eqref{fourier6} we obtain the Fourier expansion for the probing function in \eqref{probinggeneral}:
\beqn
\eta_x(y) = \f{\p w_x}{\p \nu}(y) =  \f{1}{2\pi} \sum_{n \in \mathbb{Z} } \frac{J_{n}(i \sqrt{\mu_0} r_x )}{J_n(i\sqrt{\mu_0})} e^{i n (\theta_x - \theta_y)}, \quad y \in \mathbb{S}^1  \,.
\label{coeff10}
\eqn
From the fact that
\beqn
J_m(t)\bigg\slash\frac{1}{\sqrt{2\pi|m|}}\left(\frac{et}{2|m|}\right)^{|m|}\to 1\quad \text{as}\quad m\to\infty
\eqn
for all $t>0$, we have
\beqn
\frac{J_m(i\sqrt{\mu_0}r_x)}{J_m(i\sqrt{\mu_0})}\bigg\slash r_{x}^{|m|}\to 1\quad \text{as}\quad m\to\infty.
\eqn
Hence, we observe that the coefficient in the series decays exponentially. Therefore we may approximate the probing function $\eta_x$ effectively by truncating the series \eqref{coeff10}.  This gives us an efficient method to calculate the probing function
for $\mu_0 \neq 0$, which is very useful for practical purpose.

Next, we shall calculate the kernel $K(x,z)$ for $x,z\in B_1(0)$ in \eqref{eqn:Ksub}.
For this, we first compute the Fourier expansions of $G_z$ defined as in \eqref{Green} and $\eta_x$.
For a given $g$, consider the Neumann problem:
\beqn
- \Delta v  +\mu_0 v = 0 \quad \text{ in } \quad B_1(0) \,; \quad
\f{\p v}{\p \nu} = g \quad \text{ on } \quad  \mathbb{S}^1 \,.
\eqn
Again, from the Green's representation formula, the function $v$ can be expressed in the following form
\beqn
v (z) =  \int_{\mathbb{S}^1} g (y) G_z (y) d \sigma(y) \,, \quad y \in \mathbb{S}^1  \,.
    \label{fourier7}
\eqn
On the other hand, we can calculate the Fourier expansion of function $v$ by a separation of variables:
{\small\beqn
    v (z) =  \int_{0}^{2\pi} g(\theta_y) \left( - \f{J_{0}(i \sqrt{\mu_0} r_z )}{ i \pi \sqrt{\mu_0} J_{1}(i \sqrt{\mu_0} ) } + \sum_{n \in \mathbb{Z} \backslash \{0\}}   \f{1}{i \pi \sqrt{\mu_0}} \f{J_{|n|}(i \sqrt{\mu_0} r_z )}{J_{|n|-1}(i \sqrt{\mu_0} ) - J_{|n|+1}(i \sqrt{\mu_0} ) }   e^{ i n ( \theta_z -\theta_y)} \right) d \theta_y  \,,
    \label{fourier8}
\eqn
}where $z = (r_z,\theta_z)$ is in the polar coordinate.  Comparing \eqref{fourier8} with \eqref{fourier7}, we obtain
\beqn
    G_z(y) =  - \f{J_{0}(i \sqrt{\mu_0} r_z )}{ i \pi \sqrt{\mu_0} J_{1}(i \sqrt{\mu_0} ) } + \sum_{n \in \mathbb{Z} \backslash \{0\}}   \f{1}{i \pi \sqrt{\mu_0}} \f{J_{|n|}(i \sqrt{\mu_0} r_z )}{J_{|n|-1}(i \sqrt{\mu_0} ) - J_{|n|+1}(i \sqrt{\mu_0} ) }   e^{ i n ( \theta_z -\theta_y)} \,, \quad y \in \mathbb{S}^1  \,.
    \label{coeff11}
\eqn
Then it follows from \eqref{coeff10} and \eqref{coeff11} that
\beqn
[\mathfrak{F} ( G_z ) ](n) &=&  \f{1}{i \pi \sqrt{\mu_0}} \f{J_{|n|}(i \sqrt{\mu_0} r_z )}{J_{|n|-1}(i \sqrt{\mu_0} ) - J_{|n|+1}(i \sqrt{\mu_0} ) }  e^{-  i n  \theta_z } \quad \forall \;\;n \in \mathbb{Z}\backslash\{0\}\, ,\\[2mm]
\; [\mathfrak{F} ( G_z ) ](0) &=&  - \f{J_{0}(i \sqrt{\mu_0} r_z )}{ i \pi \sqrt{\mu_0} J_{1}(i \sqrt{\mu_0} ) } \,, \label{coeffsln11} \\[2mm]
 [\mathfrak{F} ( \eta_x ) ](n) &=& \f{1}{2\pi} \frac{J_{n}(i \sqrt{\mu_0} r_x )}{J_n(i\sqrt{\mu_0})} e^{- i n \theta_x } \,  \quad \forall\;\; n \in \mathbb{Z}\,. \label{coeffsln12}
\eqn
Substituting the above expressions into \eqref{innerproducth1}, we have the following duality product
for all $x,z\in B_1(0)$,
\beqn
\langle \eta_x, G_z  \rangle_1 &=&  2\pi \sum_{n \in \mathbb{Z}} n^2 \overline{  [\mathfrak{F} ( \eta_x ) ](n)} [\mathfrak{F} ( G_z ) ](n) \notag\\
&=&   \f{1}{i \pi \sqrt{\mu_0}} \sum_{n \in \mathbb{Z} \backslash \{0\}} n^2  \f{J_{n}(i \sqrt{\mu_0} r_z ) \overline{J_{n}(i \sqrt{\mu_0} r_x )} }{\left(J_{n-1}(i \sqrt{\mu_0} ) - J_{n+1}(i \sqrt{\mu_0} ) \right) \overline{J_n(i\sqrt{\mu_0})}}  e^{i n (\theta_x - \theta_z)} \,,
\eqn
whereas the $L^2$ norm and $H^1$ semi-norm of $\eta_x$ can be obtained similarly as above.

\subsection{The rectangular domain}
In this subsection, we turn our attention to the case when $\Omega$ is a rectangular domain $(0,h)\times (-L,L)$ in $\mathbb{R}^2$ for some $h,L \in \mathbb{R}$ and calculate the probing functions introduced in \eqref{probinggeneral} and the kernel in \eqref{eqn:Ksub}. We consider the case of partial measurements along $\Gamma = \{0,h\}\times (-L,L) \subset \p \Omega$ with
$\mu_0=0$.  We know the fundamental solution $\Phi_x$
for $x = (x_1,x_2) \in \Omega$,
satisfying \eqref{phiphi}-\eqref{phidecay} with $\mu_0=0$, is given by
\beqn
\Phi_x(y) =  - \f{1}{2 \pi} \log |y - x| \,, \quad y \in \Omega \,.
\label{fundtd}
\eqn
Then we obtain from \eqref{wx} and the reflection principle that the function $w_x$ is of the following form
\beqn
    w(x,y) = - \f{1}{2 \pi}\sum_{k = 1,2} \sum_{I \in \mathbb{Z}^2} \log|y - x^{(k)}_I| + \f{1}{2 \pi} \sum_{k = 3,4} \sum_{I \in \mathbb{Z}^2} \log|y - x^{(k)}_I| \,, \quad y \in \Omega \,,
    \label{repwlong}
\eqn
where $x^{(k)}_I$ is defined, for $k = 1,2,3,4$ and $I = (i_1 ,i_2) \in \mathbb{Z}^2$, as follows:
\beqnx
\begin{aligned}
     x^{(1)}_I & = (x_1 + 2 h \, i_1 , \, x_2 +  4 L\, i_2 )\, ; \quad & x^{(2)}_I & = (x_1 + 2 h \, i_1 , \, 2L - x_2 +  4 L\, i_2 ) \,; \\
     x^{(3)}_I & = (- x_1 + 2 h \, i_1 , \, x_2 +  4 L\, i_2 )\, ; \quad & x^{(4)}_I & = (- x_1 + 2 h \, i_1 , \, 2L - x_2 +  4 L\, i_2 ) \,.
\end{aligned}
\eqnx
Therefore, writing $x^{(k)}_I = ( x^{(k)}_{I,1}, x^{(k)}_{I,2} )$ for $k = 1,2,3,4$ and $I = (i_1 ,i_2) \in \mathbb{Z}^2$, we have the following series expansion for the probing function defined as in \eqref{probinggeneral} for all $y \in  \{h\}\times (-L,L) \subset \Gamma$:
\beqn
\eta_x(y) = \f{\p w_x}{\p y_1}(y)
=
- \f{1}{2 \pi} \left(
\sum_{k = 1,2} \sum_{I \in \mathbb{Z}^2} \f{h - x^{(k)}_{I,1}}{|(h,y_2) - x^{(k)}_I|^2} - \sum_{k = 3,4} \sum_{I \in \mathbb{Z}^2} \f{h - x^{(k)}_{I,1}}{|(h, y_2) - x^{(k)}_I|^2} \right) \, ,
\eqn
whereas for $y \in  \{0\}\times (-L,L) \subset \Gamma$, with a similar argument, we have
\beqn
\eta_x(y) = - \f{\p w_x}{\p y_1}(y) =
- \f{1}{2 \pi} \left(
\sum_{k = 1,2} \sum_{I \in \mathbb{Z}^2} \f{x^{(k)}_{I,1}}{|(0,y_2) - x^{(k)}_I|^2} - \sum_{k = 3,4} \sum_{I \in \mathbb{Z}^2} \f{x^{(k)}_{I,1}}{|(0, y_2) - x^{(k)}_I|^2} \right) \,.
\eqn
This gives an explicit expression of the probing function $\eta_x$ for all $x \in \Omega$.
But the above series representation of $\eta_x$ is tedious to work with if we intend to calculate the kernel $K$ defined in \eqref{eqn:Ksub}. However, we can substitute \eqref{fundtd} into \eqref{repsmooth} to obtain the following representation
of $w_x$ for $x \in \Omega$:
\beqnx
    w(x,y) &=&  - \f{1}{2 \pi}\log|x - y| - \psi(x,y) \,, \quad x,y \in \Omega \,,
\eqnx
where $\psi(x,y) := \psi_x(y)$ is defined as in \eqref{psismooth} and is a smooth function with respect to $x,y \in \Omega$.
Then we can derive the following expression of the probing function for $x \in (0,h)\times (-L,L)$:
\beqn
\eta_x(y) &=& \f{\p w_x}{\p y_1}(y) =  - \f{1}{2 \pi}\f{h-x_1}{|(h,y_2) - x|^2} - \f{\p \psi}{\p y_1} (x,y) \,, \quad y \in \{h\}\times (-L,L) \,,
\label{etaeta1}\\
\eta_x(y) &=&- \f{\p w_x}{\p y_1}(y) =  - \f{1}{2 \pi}\f{x_1}{|(0,y_2) - x|^2} + \f{\p \psi}{\p y_1} (x,y) \,, \quad y \in \{0\}\times (-L,L) \,.
\label{etaeta2}
\eqn
On the other hand, subtracting \eqref{Greenhomozero} from \eqref{phiphi} and substituting \eqref{fundtd} into it, we can obtain the following representation of $G_z$ for $z\in \Omega$ based on the fundamental solution $\Phi_z$:
\beqn
G_z(y)=\Phi_z(y)-\chi (z,y) = - \f{1}{2 \pi}\log|z - y| -\chi (z,y) \, ,
\label{hahasmooth}
\eqn
where $\chi (z,\cdot) $ is a function satisfying the following system:
\beqn
-\Delta\chi (z,\cdot)=0 \text{ in }\Omega\,; \quad
\frac{\partial \chi (z,\cdot) }{\partial \nu}=\frac{\partial \Phi_z}{\partial\nu} - \f{1}{|\Gamma|}\quad \text{ on } \quad\p \Om\,; \quad
\quad \int_{\Gamma} \chi (z,\cdot) \, d \sigma  = \int_{\Gamma} \Phi_z  \, d \sigma \,,
\label{chismooth}
\eqn
and is a smooth function with respect to $z,y \in \Omega$. From \eqref{etaeta1}-\eqref{hahasmooth} and the Green's identity, we can see that the $\langle \, \cdot \, , \, \cdot \, \rangle_1$ duality product of $\eta_x$ and $G_z$ can be expressed in the following form for all $x, z \in \Omega$:
\beqn
&&
\langle \, \eta_x \, , \, G_z \, \rangle_1 = \textbf{(I)} + \textbf{(II)}  \label{ans11}
\eqn
where the terms $\textbf{(I)}$ and $\textbf{(II)}$ as as follows:
\beqn
&&
\textbf{(I)} \notag\\
&=&
 \int_{-L}^L \f{\p }{\p y_2} \left( - \f{1}{2 \pi}\f{h-x_1}{|(h,y_2) - x|^2} - \f{\p \psi}{\p y_1} \Big(x,(h,y_2)\Big)  \right)
 \f{\p }{\p y_2} \left( - \f{1}{2 \pi}\log|(h,y_2) - z| -\chi \Big(z,(h,y_2)\Big) \right) \, d y_2 \notag\\
&& + \int_{-L}^L
 \f{\p }{\p y_2} \left( - \f{1}{2 \pi}\f{x_1}{|(0,y_2) - x|^2} + \f{\p \psi}{\p y_1} \Big(x,(0,y_2)\Big)  \right)
\f{\p }{\p y_2} \left( - \f{1}{2 \pi}\log|(0,y_2) - z| -\chi \Big(z,(0,y_2)\Big) \right)
 \, d y_2 \notag\\
&=&
 \int_{-L}^L  \left( \f{1}{\pi}\f{(h-x_1)(y_2-x_2)}{|(h,y_2) - x|^4} - \f{\p^2 \psi}{\p y_1 y_2} \Big(x,(h,y_2)\Big)  \right)
 \left( - \f{1}{2 \pi} \f{y_2-z_2}{|(h,y_2) - z|^2} -\f{\p \chi }{\p y_2}  \Big(z,(h,y_2)\Big) \right) \, d y_2 \notag\\
&& + \int_{-L}^L
 \left( - \f{1}{\pi}\f{x_1(y_2-x_2)}{|(0,y_2) - x|^4} + \f{\p^2 \psi}{\p y_1 y_2} \Big(x,(0,y_2)\Big)  \right)
 \left( - \f{1}{2 \pi}\f{y_2-z_2}{|(0,y_2) - z|^2} -\f{\p \chi }{\p y_2} \Big (z,(0,y_2)\Big) \right)
 \, d y_2 \notag\\
 &=&
  C_1 \int_{-L}^L  \left( \f{(h-x_1)(y_2-x_2)(y_2-z_2)}{|(h,y_2) - x|^4 |(h,y_2) - z|^2} + \f{x_1(y_2-x_2)(y_2-z_2)}{|(0,y_2) - x|^4|(0,y_2) - x|^2} + \varepsilon_1 (x,z,y_2)
 \right)\, d y_2\,, \label{ans11i}
\eqn
where $C_1$ is a constant and $\varepsilon_1 (x,z,y_2)$ is a function whose order of magnitude is smaller than the first two terms in the integrand; whereas with the help of \eqref{zero110}, the second term can be calculated directly as
\beqn
\textbf{(II)} = \left[ \f{\p \eta_x }{ \p x_2 } (h, y_2)
  G_z (h, y_2) \right]_{y_2 = -L}^L   - \left[ \f{\p \eta_x }{ \p x_2 } (0, y_2)
  G_z (0, y_2) \right]_{y_2 = -L}^L = 0
 \,. \label{ans11ii}
\eqn
Using the same technique, the $L^2$ norm and $H^1$ semi-norm of $\eta_x$ can be expressed as follows:
\beqn
    |\eta_x|^2_{H^0(\Gamma)} = C_2 \int_{-L}^L \f{(h-x_1)^2}{|(h,y_2) - x|^4} + \f{x_1^2}{|(0,y_2) - x|^4} + \varepsilon_2 (x, y_2) \, d y_2 \,, \label{ans12}
\eqn
while
\beqn
|\eta_x|^2_{H^1(\Gamma)}  = C_3 \int_{-L}^L \f{(h-x_1)^2(y_2-x_2)^2}{|(h,y_2) - x|^8} + \f{x_1^2(y_2-x_2)^2}{|(0,y_2) - x|^8} + \varepsilon_3 (x, y_2) \, d y_2 \label{ans13}\,,
\eqn
where, for $i = 2,3$, $\varepsilon_i (x,y_2)$ are some functions with its order of magnitude being
smaller than the first two terms of the integrand, and $C_i$ are some constants.
Substituting \eqref{ans11}-\eqref{ans13} into \eqref{eqn:Ksub}, we can see that the kernel $K(x,z)$ is relatively large when $x \approx z$, while it is relatively small outside.
Figure \ref{kernalori} shows the values of the kernel $K(x,z)$ with $z = (0.220, -0.307)$ for $x\in (0,h)\times (-L,L)$. This supports us to use the index function defined in \eqref{eqn:inner} for locating inhomogeneities in the rectangular domain for the DOT problem.
\begin{figurehere}
     \begin{center}
     \vskip -0.18truecm
           \scalebox{0.5}{\includegraphics{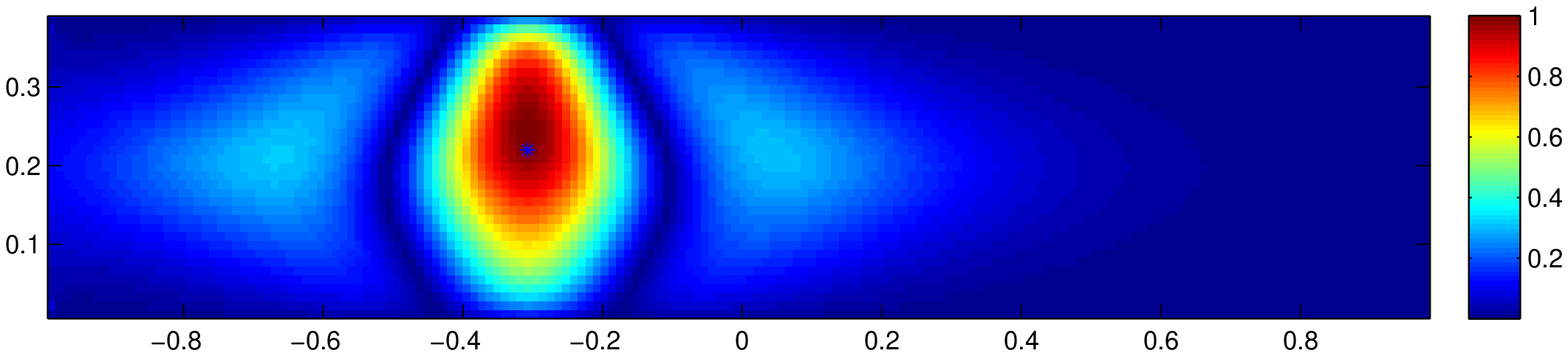}}\\
            \vskip -0.4truecm
     \caption{\small Kernel $K(\,\cdot\,,z)$ on a rectangular domain with $z = (0.220, -0.307)$ marked as a blue star. } \label{kernalori}
     \end{center}
 \end{figurehere}

\section{Alternative characterization of index functions and scaling effects} \label{sec:scale}

In this section we shall present an alternative characterization of the index function defined in \eqref{eqn:inner}
for the following two special cases, or two most frequently used sampling domains:
$\Om$ is an arbitrary open connected domain with $\Gamma$ being a closed surface
on $\p\Omega$;
and $\Omega = (0,h)\times (-L,L)$ is a rectangular domain with $\Gamma = \{0,h\}\times (-L,L)$. With these
characterizations, we will introduce an additional scaling function which will greatly improve the imaging of the index function as well as essentially reducing the computational effort for the index function.

We first define an auxiliary function which will be very useful to represent the index function $I$.
Let $\varphi$ be a function which satisfies the following homogeneous diffusion system: 
\beqn
-\Delta\varphi+\mu_0\varphi=0 \quad\text{ in }\quad\Omega\,; \quad
\varphi = \Delta_\Gamma (f-f_0) \quad\text{ on } \quad\Gamma; \quad
\frac{\partial \varphi }{\partial\nu}&=0\quad \text{ on } \quad\partial\Omega\backslash\Gamma\,.
\label{defphi}
\eqn
Then from the definition of the duality product $\langle \, \cdot \, , \, \cdot \, \rangle_1$ and the 
Green's identity on $\Gamma$, we have
\beqn
 \langle \eta_x, f - f_0  \rangle_1 =  - \int_{\Gamma} \eta_x \Delta_{\Gamma} (f - f_0) \, d \sigma_{\Gamma} -  \int_{\p \Gamma} (f - f_0) \f{\p \eta_x}{ \p n } \, d \sigma_{\p \Gamma}
 + \int_{\p \Gamma}  \eta_x \f{\p }{ \p n }(f - f_0) \, d \sigma_{\p \Gamma}\,
\label{varphirep}
\eqn
for any $x \in \Omega$,
where $d \sigma_{\Gamma}$ and $d \sigma_{\p \Gamma}$ represent the respective volume
and surface elements on $\Gamma$ and $\p \Gamma$ while the vector $n$ is the unit vector field normal to $\p \Gamma$.

When $\Om$ is an arbitrary open connected domain but $\Gamma$ is a closed surface,
we know $\p \Gamma = \emptyset$.
Then it follows from \eqref{wx}, \eqref{probinggeneral}, \eqref{defphi}-\eqref{varphirep} and the Green's formula that
\beqn
 \langle \eta_x, f - f_0  \rangle_{1} &=&  - \int_{\Gamma} \eta_x  \Delta_{\Gamma} (f - f_0) \, d \sigma_{\Gamma}  
= - \int_{\Gamma} \varphi \frac{\partial w_x}{\partial\nu}   \, d \sigma_{\Gamma}  \notag \\
&=& \int_{\partial\Omega} \left( w_x \frac{\partial \varphi }{\partial\nu}- \varphi \frac{\partial w_x}{\partial\nu}    \right)\, d \sigma
= \int_{\Omega} \left( w_x \Delta \varphi - \varphi  \Delta w_x \right)  \, d y
=\varphi(x).
\label{representphi1}
\eqn
On the other hand, when $\Omega$ is a rectangular domain, $\Omega = (0,h)\times (-L,L)$,
and $\Gamma = \{0,h\}\times (-L,L)$, we have from \eqref{defphi}-\eqref{varphirep} that
\beqn
 \langle \eta_x, f - f_0  \rangle_{1} &=& - \int_{\Gamma} \eta_x \Delta_{\Gamma} (f - f_0) \, d \sigma_{\Gamma} 
 \notag \\
& &  -  (u - u_0) (0,L) \f{\p \eta_x }{ \p x_2 } (0,L) + (u - u_0) (0,-L) \f{\p \eta_x }{ \p x_2 } (0,-L) \notag\\
& & -   (u - u_0) (h,L)  \f{\p \eta_x }{ \p x_2 } (h,L) + (u - u_0) (h,-L) \f{\p \eta_x }{ \p x_2 } (h,-L) \notag\\
& &  +  \eta_x(0,L) \f{\p (u - u_0)}{ \p x_2 } (0,L) - \eta_x(0,-L) \f{\p (u - u_0)}{ \p x_2 } (0,-L) \notag\\
& & +  \eta_x(h,L) \f{\p (u - u_0)}{ \p x_2 } (h,L) - \eta_x(h,-L) \f{\p (u - u_0)}{ \p x_2 } (h,-L)\,.
\label{intermediate}
\eqn
However, substituting
\eqref{zero110} and
\eqref{zero11} into the above expression, we can directly see that all the boundary terms in the expression are zero,
hence we arrive at the following from \eqref{wx}, \eqref{probinggeneral} and the Green's formula that
\beqn
 \langle \eta_x, f - f_0  \rangle_{1} = - \int_{\Gamma} \eta_x \Delta_{\Gamma} (f - f_0) \, d \sigma_{\Gamma} = - \int_{\Gamma} \frac{\partial w_x}{\partial\nu}  \varphi \, d \sigma_{\Gamma}
=\int_{\partial\Omega} \left(w_x \frac{\partial \varphi }{\partial\nu}- \varphi\frac{\partial w_x}{\partial\nu} \right) \, d \sigma = \varphi(x).
\label{representphi2}
\eqn
Therefore, we can see from both \eqref{representphi1} and \eqref{representphi2} that, in both cases we consider, we have the following representation of the  duality product $ \langle \, \cdot \, , \, \cdot \,  \rangle_{1}$ of $\eta_x$ and $f-f_0$:
\beqn
 \langle \eta_x, f - f_0  \rangle_{1} = \varphi(x).
\label{representphi}
\eqn
Substituting this expression into \eqref{eqn:inner}, we derive the following formula for the index function
\beqn
I(x)=
\f{ \langle \eta_x, f - f_0  \rangle_{1} }{|\eta_x|^{\f{1}{2}}_{H^1(\Gamma)}|\eta_x|^{\f{3}{4}}_{H^0(\Gamma)}}
=
\f{ \varphi(x) }{|\eta_x|^{\f{1}{2}}_{H^1(\Gamma)}|\eta_x|^{\f{3}{4}}_{H^0(\Gamma)}}
\,, \quad \,x \in \Omega \,.
\label{eqn:inner:rep}
\eqn
For this expression, we suggest an appropriate scaling function
\beqn
S(x)=\frac{1}{||\varphi||_{L^\infty(\Omega)} +|\varphi(x)|}\,, \quad x \in \Omega \,.
\eqn
One may note from the calculations similar to \eqref{ans12}-\eqref{ans13} that
the $L^2(\Gamma)$ norm and $H^1(\Gamma)$ semi-norm of $\eta_x$ can be actually approximated by those of the fundamental solution $\Phi_x$ defined as in \eqref{phiphi}-\eqref{phidecay}, namely
\beqn
|\eta_x|_{H^1(\Gamma)} \approx \left|\f{\p \Phi_x}{\p \nu}\right|_{H^1(\Gamma)} \,,   \quad |\eta_x|_{L^2(\Gamma)} \approx \left|\f{\p \Phi_x}{\p \nu}\right|_{L^2(\Gamma)} \, , \quad  \forall x \in \Omega\,.
\eqn
With these approximations, we propose the following modified version of the index function in \eqref{eqn:inner:rep}:
\beqn
\widetilde{I}(x) := \f{ 1 }{\left|\f{\p \Phi_x}{\p \nu}\right|^{\f{1}{2}}_{H^1(\Gamma)}\left|\f{\p \Phi_x}{\p \nu}\right|^{\f{3}{4}}_{H^0(\Gamma)}} \frac{\varphi(x)}{||\varphi||_{L^\infty(\Omega)} +|\varphi(x)|} \,, \quad x \in \Omega \,.
\label{eqn:inner:new}
\eqn
The numerical experiments in the next section will confirm that this modified index function improves
the images in locating inhomogeneities in the DOT problem.

In what follows, we would like to give a comparison between the original index function $I$ and the modified index function $\widetilde{I}$. For this purpose, given $z \in \Omega$, let $\mu^z$ be an absorption coefficient of the following form
\beqn
\mu^z(x) = \mu_0 + \mu_1 \chi_{B_{\varepsilon}(z)}(x) \, , \quad x \in \Omega\,,
\eqn
where $\varepsilon \in \mathbb{R}$ is a very small number, $ \mu_1 \in \mathbb{R}$ is a non-zero constant, $B_{\varepsilon}(z)$ is an open ball of radius $\varepsilon$ and center $z$, and $\chi_{B_{\varepsilon}(z)}$ is its characteristic function.  Then from \eqref{eqn:property_b} and  \eqref{eqn:property_b_homozero}, we can see that, if $\mu_0 \neq 0$, the scattered potential $f$ along $\Gamma$ satisfies
\beqn
(f-f_0)(\xi)\approx a_z G_{z}(\xi) \, , \quad \xi \in \p \Om\,,
\eqn
for some $a_z \in \mathbb{R}$; whereas if $\mu_0 = 0$, we have
\beqn
(f-f_0)(\xi)\approx a_z G_{z}(\xi) - \f{1}{|\p\Om|} \int_{\p\Om} f d \sigma \, , \quad \xi \in \p \Om\,.
\eqn
In either case, we can choose a value of $\mu_1$ such that $a_z = 1$.  Then, from \eqref{indexexpression} and \eqref{indexexpression2}, we can see that
\beqn
 I(x) \approx a_z K(x,z) = K(x,z) \,, \quad x \in \Omega \,,
\eqn
where $K$ is the function defined as in (\ref{eqn11:K}). Now, given $z\in \Omega$, with this choice of $\mu^z$, we can also calculate the modified index function $\widetilde{I}(x)$ in \eqref{eqn:inner:new} from the the scattered potential $f$ along $\Gamma$.
We denote this index function $\widetilde{I}(x)$ as $\widetilde{K}(\,\cdot \,,z)$, i.e.,
\beqn
 \widetilde{K}(x,z) := \widetilde{I}(x) \,, \quad x \in \Omega \,.
\eqn
We can now compare the behaviors of the two functions $K(\,\cdot\,,z)$ and $\widetilde{K}(\,\cdot\,,z)$ to get a better understanding of the two index functions and compare their performances.   Figure \ref{kernalmod} shows the values of the function $\widetilde{K}(x,z)$ with $z = (0.220, -0.307)$ for $x\in (0,h)\times (-L,L)$.
\begin{figurehere}
     \begin{center}
     \vskip -0.18truecm
           \scalebox{0.5}{\includegraphics{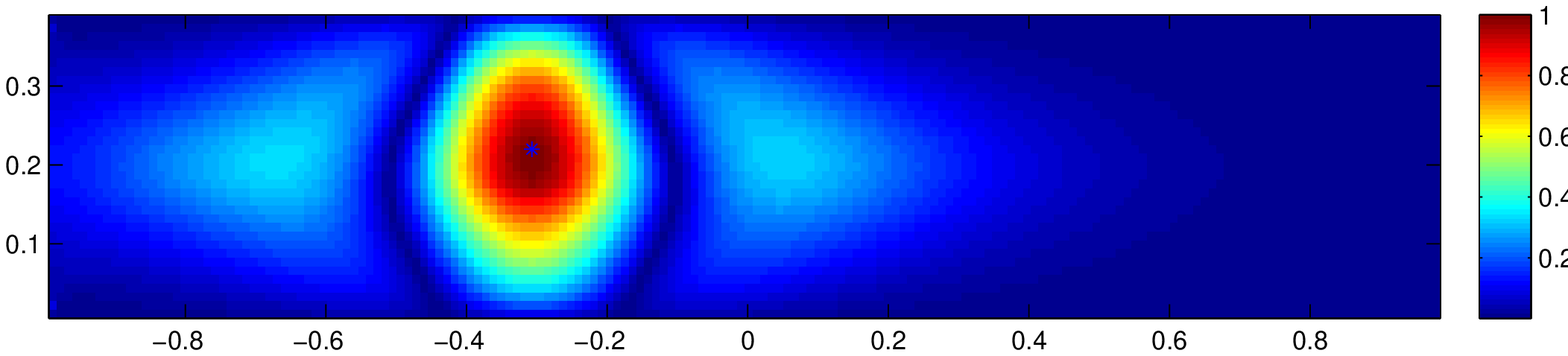}}\\
            \vskip -0.4truecm
     \caption{\small The function $\widetilde{K}(\,\cdot\,,z)$ on a rectangular domain with $z = (0.220, -0.307)$ marked by a blue star.} \label{kernalmod}
     \end{center}
 \end{figurehere}
Comparing Figure \ref{kernalmod} with Figure \ref{kernalori},
we observe the maximum point 
of the function $K(\,\cdot\,,z)$ in Figure \ref{kernalori} is shifted a bit upward and away from the point $z$,
whereas the maximum point 
of $\widetilde{K}(\,\cdot\,,z)$ in Figure \ref{kernalmod} matches perfectly 
with the point $z$.
This phenomenon is of crucial importance, and accounts for the better performance and higher accuracy of this modified index in \eqref{eqn:inner:new} to locate inhomogeneities for the DOT problem.


Moreover, we would like to emphasize that the evaluation of the modified index function $\widetilde{I}(x)$ defined as in \eqref{eqn:inner:new} is computationally much cheaper than \eqref{eqn:inner}.
The function $\varphi$ in the expression \eqref{eqn:inner:new} can be found by solving a forward elliptic
system \eqref{defphi} once a priori for which fast solvers with nearly optimal complexity
are available for general domains, such as multigrid methods and domain decomposition methods,
while the semi-norms of the fundamental solution $\Phi_x$ can be efficiently approximated by numerical quadrature rules since the closed form of $\Phi_x$ are explicitly known.

%
Therefore this modified index function $\widehat{I}$ shall serve as a very efficient and robust method to locate inhomogeneities for the DOT problem.

\section{Numerical examples} \label{sec:num}
In this section, we shall present several numerical examples to illustrate the effectiveness
of the newly proposed DSM method for the reconstruction of inhomogeneous inclusions in the DOT problem.

We consider the sampling region $\Omega = (-1,1) \times (0,0.4) $, and the absorption coefficient for homogeneous background $\mu_0 = 0$ in $\Omega$.
In each of the following examples, there are some inhomogeneous inclusions placed inside $\Omega$, with their absorption coefficients always set to be $\mu= 50$.
Our choice of $\Gamma$ is $(-1,1) \times \{0,0.4\}$.
In order to collect our observed data of the forward problem, we solve \eqref{eqn11} by second order central
finite difference method
with a fine mesh of size $0.004$ and the boundary flux $g = 1$ on $ (-1,1) \times \{0.4\}$, $g = -1$ on $ (-1,1) \times \{0\}$ and $g = 0$ on $\p \Om \backslash \Gamma$.
The boundary flux $g$ is chosen as above so as to create a uniform flow from the bottom to the top of the domain.
We have observed from our numerical experiments
that this choice of boundary flux is very effective in obtaining reasonable Cauchy data for our concerned imaging.
The scattered potential $f_s:= f - f_0$ is then measured along $\Gamma$.
We would like to emphasize that, in each of our following examples, we only collect the scattered potential from a single choice of boundary flux, therefore only one pair of the Cauchy data is available for our reconstruction. So the resulting
inverse problem is a severely ill-posed problem.
In order to test the robustness of our reconstruction algorithm, we introduce some random noise in the scattered potential as follows:
\begin{equation}
f^\delta_s(x) = f_s(x) + \varepsilon \delta \max_{x} |f_s(x)| \,, \label{noise}
\end{equation}
where $\delta$ is uniformly distributed between $-1$ and $1$ and $\varepsilon$ corresponds to the noise level in the data, which is always set to be $\varepsilon = 5 \%$ in all our examples.

From the noisy observed data $f^\delta_s$, we then use the DSM method to solve the DOT problem by calculating the index function $I$ introduced in \eqref{eqn:inner} and the modified index function $\widetilde{I}$ in \eqref{eqn:inner:new}.
From our numerical experiments, we observe that the square of modified index $\widetilde{I}^2$ gives us sharper images of the inclusions, hence providing a more accurate estimate of the support of the inhomogeneities $D$.  Therefore, in each of the following examples, we provide the reconstructed images from all the three indices $I$, $\widetilde{I}$ and $\widetilde{I}^2$. 
For the better illustrative purpose and comparison, we normalize each of the aforementioned indices with a normalizing constant such that the maximum of each index is $1$.
The mesh size in our reconstruction process is chosen to be $0.011$.

\begin{figurehere}
     \begin{center}
     \vskip -0.25truecm
           \scalebox{0.4}{\includegraphics{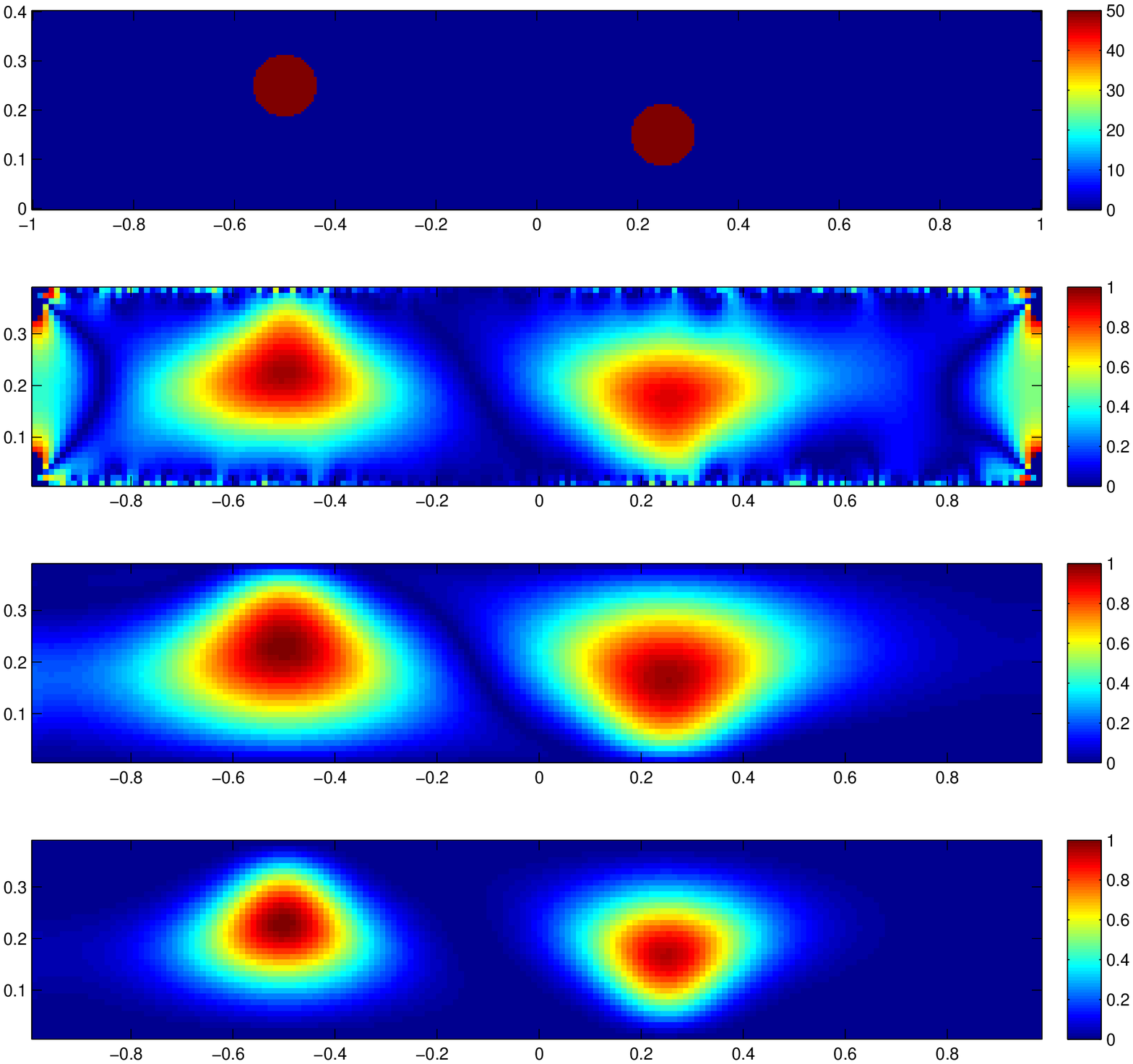}}\\
            \vskip -0.8truecm
     \caption{\small From top to bottom: exact medium in Example 1; index function $I$; modified index function $\widetilde{I}$; square of modified index function $\widetilde{I}$.}\label{test1}
     \end{center}
 \end{figurehere}

\textbf{Example 1}. This example tests a medium with two circular inclusions of radius $0.065$, which are respectively centered at $(-0.5, 0.25)$ and $(0.25, 0.15)$; see Figure \ref{test1} (top).
Figure \ref{test1} (second to bottom) show the respective reconstructed images from the index function $I$, the modified index function $\widetilde{I}$ and the square of $\widetilde{I}$.  We can see from the figures that the recovered scatterers are well separated, and the locations of the scatterers are accurately reconstructed. We may also observe that the artifacts due to the presence of noise are effectively removed by the modified index function. Moreover, the squared modified index function 
provides the sharpest image of the inclusions and gives most accurate locations of the inclusions as well as being the most robust to noise.

\textbf{Example 2}. In this example, our medium consists of $4$ circular inclusions of radius $0.065$ with their corresponding positions: $(-0.3, 0.1)$, $(-0.3,0.3)$, $(0.3, 0.1)$ and $(0.3, 0.3)$; see Figure \ref{test2} (top).
The reconstructed images from the index $I$, the modified index $\widetilde{I}$ and its square $\widetilde{I}^2$ are shown in Figure \ref{test2} (second to bottom).
We observe that both the locations and sizes of the reconstructed scatterers agree well with those of the true ones.
We can also see that the modified index function $\widetilde{I}$ gives a better estimate of the sizes and locations of the inclusions as well as removes all the artifacts from the index function $I$.  
In addition, the square of modified index provides the best image, because it not only gives more focused and accurate shapes and locations of the inclusions, but also significantly reduce the heavy shadows from the image.
Considering the severe ill-posedness of the problem and the challenging case of 4 scatterers staying very close
to each other in the vertical direction and close to the boundaries, our reconstruction seem to be quite satisfactory with only one pair of the Cauchy data.

\begin{figurehere}
     \begin{center}
     \vskip -0.2truecm
           \scalebox{0.4}{\includegraphics{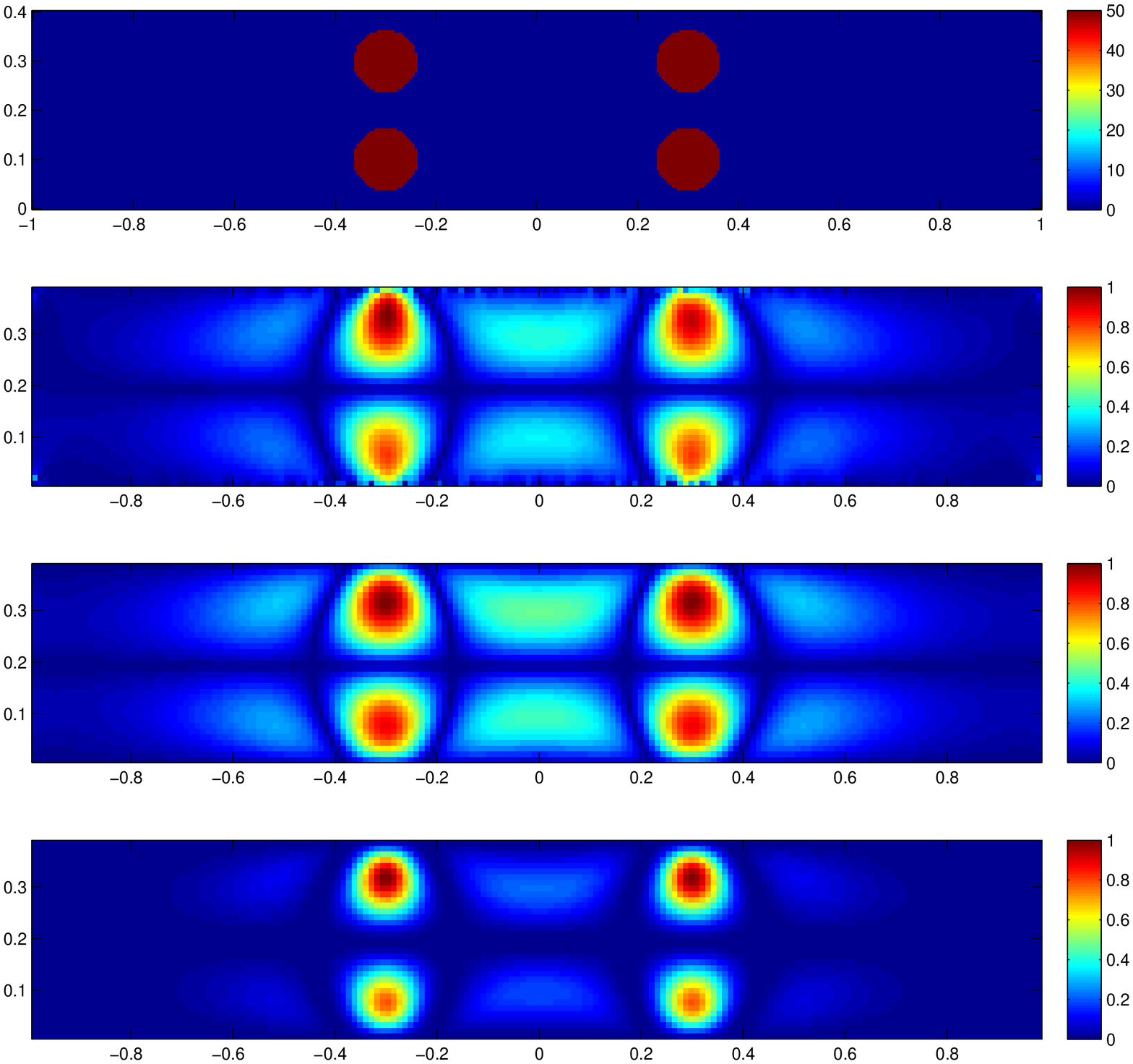}}\\
            \vskip -0.8truecm
     \caption{\small From top to bottom: exact medium in Example 2; index function $I$; modified index function $\widetilde{I}$; square of modified index function $\widetilde{I}$.}\label{test2}
     \end{center}
 \end{figurehere}

\textbf{Example 3}. A medium with $4$ circular inclusions of radius $0.065$ centered at $(-0.5, 0.3)$, $(-0.3, 0.1)$, $(0, 0.3)$ and $(0.3, 0.1)$ is investigated in this example; see Figure \ref{test3} (top).  Figure \ref{test3} (second to bottom) displays
the three reconstructed images from the index $I$, the modified index $\widetilde{I}$ and its square $\widetilde{I}^2$ respectively.
Although we observe some minor shifting of the recovered scatterers in the images, they are still well separated and located with reasonable accuracy, and the artifacts for the index function $I$ are all removed by the modified index function $\widetilde{I}$. Furthermore, the squared modified index function $\widetilde{I}^2$ gives the best reconstruction by providing the most accurate estimate of locations and shapes of the scatterers as well as removing part of the heavy shadows from $\widetilde{I}$.

\begin{figurehere}
     \begin{center}
     \vskip -0.2truecm
           \scalebox{0.4}{\includegraphics{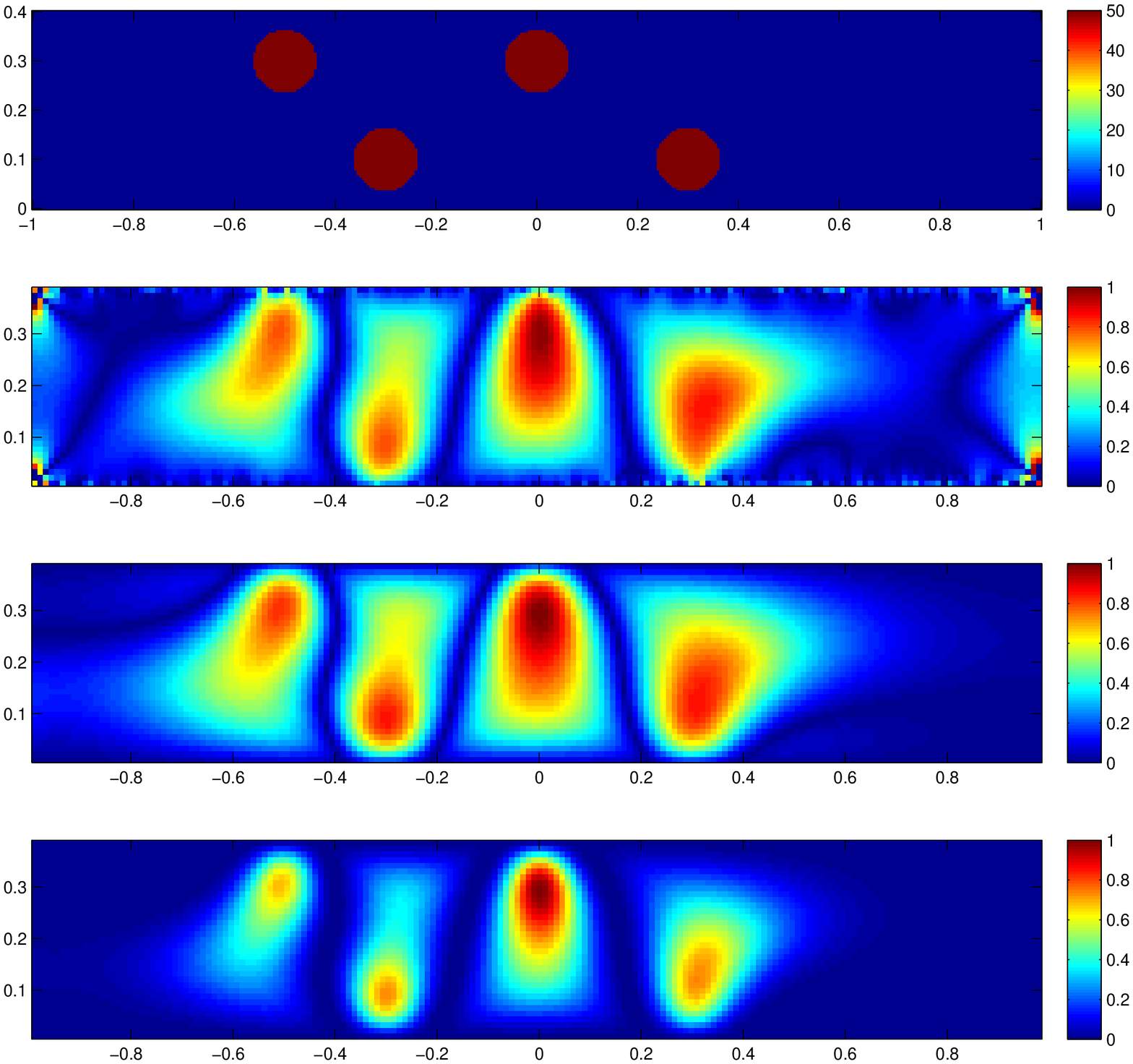}}\\
            \vskip -0.8truecm
      \caption{\small From top to bottom: exact medium in Example 3; index function $I$; modified index function $\widetilde{I}$; square of modified index function $\widetilde{I}$.}\label{test3}
     \end{center}
 \end{figurehere}



 \textbf{Example 4}. This example investigates a rectangular inclusions of width $0.1$ and length $0.2$ centered at $(0, 0.15)$; see Figure \ref{test5} (top).\
Figure \ref{test5} (second to bottom) displays
the three respective reconstructed images from $I$, $\widetilde{I}$ and $\widetilde{I}^2$.
We can observe from the reconstructed images that the reconstructed inclusion is mildly shortened and shifted downwards, and that the modified indicator function $\widetilde{I}$ removes the artifacts from the index function $I$. In addition, the squared index $\widetilde{I}^2$ significantly removes the shadows from $\widetilde{I}$, 
hence providing the most reliable reconstruction of the inclusion.

\textbf{Example 5}. In this last example, two rectangular inclusions of width $0.05$ and length $0.4$ are centered at $(-0.1, 0.125)$ and $(-0.1, 0.275)$ in $\Omega$; see Figure \ref{test6} (top).
The reconstructed images by the three indices $I$, $\widetilde{I}$ and $\widetilde{I}^2$ are respectively shown in Figure \ref{test6} (second to bottom).
Due to the ill-posed nature of the DOT, the reconstructed inclusions are not free from shadows and oscillations.
However, the overall profile stands out clearly, and both the locations and the shapes of the inclusions are well reconstructed. Furthermore, the shadows are greatly reduced by the squared modified index function $\widetilde{I}^2$.

In all of the above examples,
we may observe that the reconstructions seem to be rather satisfactory,
considering the severe ill-posedness of the inverse problem, and the facts that
the $5\%$ random noise is added in the observed data and only one pair of the Cauchy data is available.
Moreover, the square $\widetilde{I}^2$  of the modified index function provides the most accurate locations for the inclusions and is the most robust against noise. 
Considering the good accuracy of reconstruction, computational efficiency and robustness to noise, the new DSM
proves to be an effective new numerical tool for the reconstruction of the DOT problem.

\begin{figurehere}
     \begin{center}
     \vskip -0.3truecm
           \scalebox{0.4}{\includegraphics{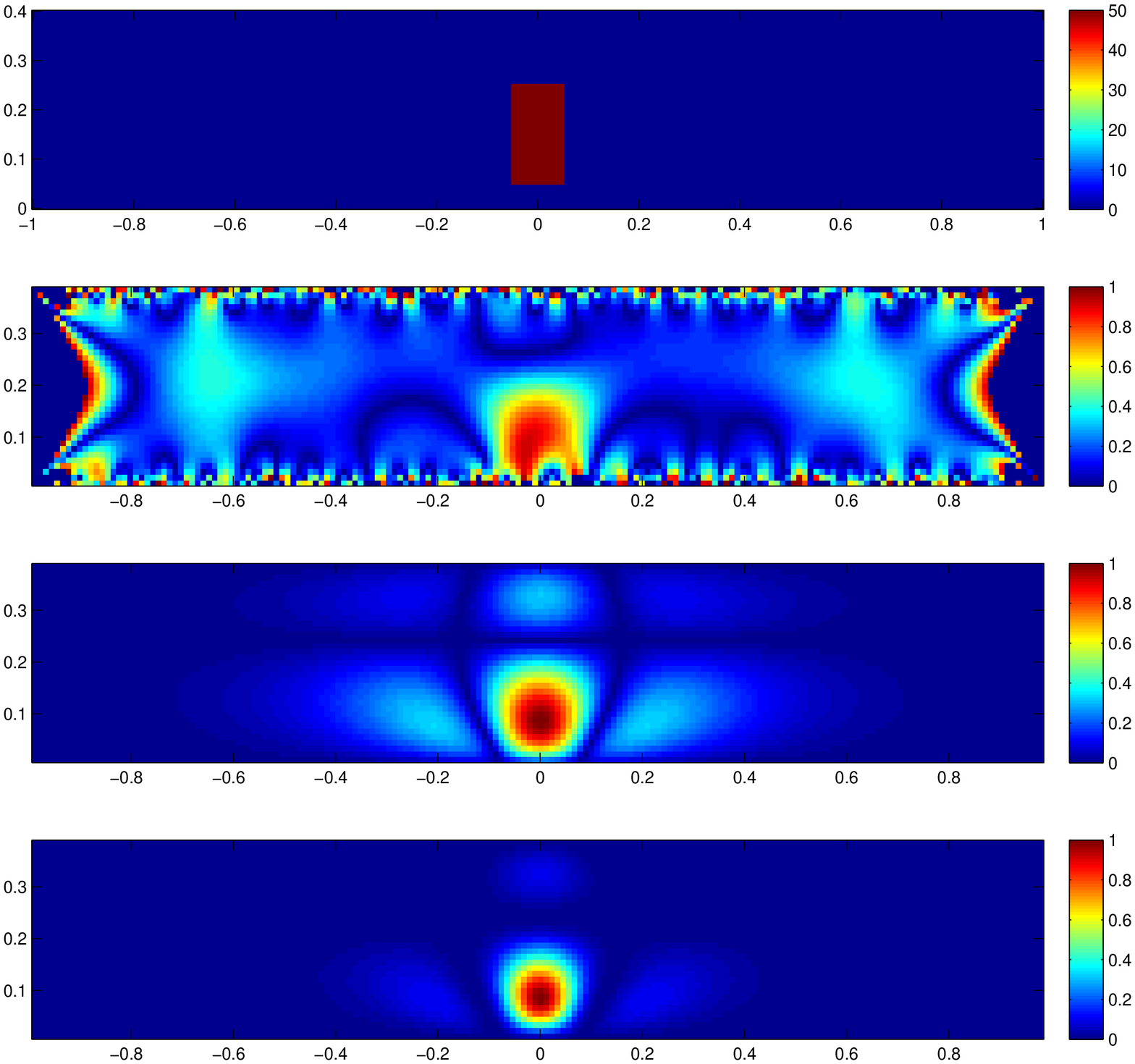}}\\
     \vskip -0.8truecm
       \caption{\small From top to bottom: exact medium in Example 4; index function $I$; modified index function $\widetilde{I}$; square of modified index function $\widetilde{I}$.}\label{test5}
      \end{center}
 \end{figurehere}

\begin{figurehere}
     \begin{center}
     \vskip -0.3truecm
           \scalebox{0.4}{\includegraphics{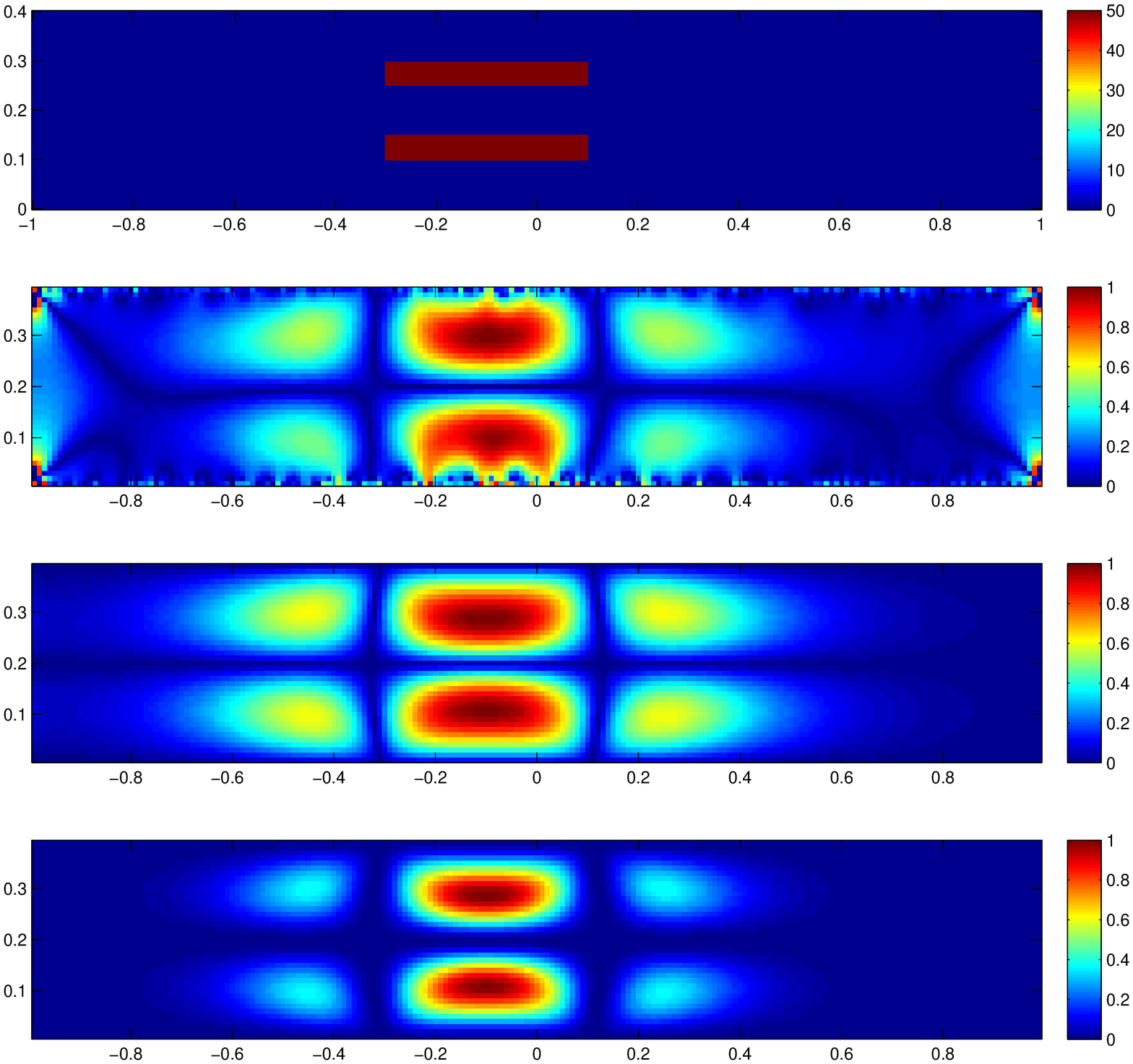}}\\
            \vskip -0.8truecm
     \caption{\small From top to bottom: exact medium in Example 5; index function $I$; modified index function $\widetilde{I}$; square of modified index function $\widetilde{I}$.}\label{test6}
     \end{center}
 \end{figurehere}

\section{Concluding remarks} \label{sec:con}

In this work we have presented a novel direct sampling method to locate inhomogeneities inside a homogeneous background for the diffusive optical tomography problem, applicable even to one scattered potential data and
in both full and limited aperture cases.
It involves computing only a well-posed elliptic system involving the scattered potential as its boundary condition, therefore
it can be viewed as a direct method.
This method is very easy to implement and computationally inexpensive since it does not require any matrix operations solving ill-posed integral equations or any optimisation process as most existing algorithms do.  In particular, numerical experiments have demonstrated the effectiveness and robustness of the proposed method.
This new method provides an efficient numerical strategy and a new promising direction for solving the DOT problem.

\end{document}